\documentclass[twocolumn, aps, longbibliography,unsortedaddress]{revtex4-1}
\usepackage[colorlinks=true,%
bookmarks=false,%
linkcolor=green,%
urlcolor=blue,%
citecolor=magenta,%
breaklinks]{hyperref}
\usepackage[english]{babel}

\usepackage[bb=px]{mathalpha}

\usepackage{graphicx}% Include figure files
\usepackage{dsfont}
\usepackage{dcolumn}% Align table columns on decimal point
\usepackage{bm}% bold math
\usepackage{color}
\usepackage[normalem]{ulem}
\usepackage{simplewick}
\usepackage{braket}
\usepackage{float}
\usepackage{multirow}
\usepackage[utf8]{inputenc}
\usepackage[caption = false]{subfig}
\usepackage{enumitem}

\usepackage{mathtools}

\usepackage{tikz}
\usetikzlibrary{quantikz}

\usepackage{algorithm}
\usepackage{algpseudocode}

\usepackage{notes2bib} %to put notes into bibliography

\begin{document}

\title{From Vlasov-Poisson to Schr\"odinger-Poisson: dark matter simulation with a quantum variational time evolution algorithm}

\author{Luca Cappelli}
\affiliation{Dipartimento di Fisica dell'Universit\`a di Trieste \\
IBM Quantum, IBM Research – Zurich, 8803 Rüschlikon, Switzerland \\
INAF - Osservatorio Astronomico di Trieste, via Tiepolo 11, I-34131, Trieste, Italy\\
ICSC - Italian Research Center on High Performance Computing, Big Data and Quantum Computing
}
\author{Francesco Tacchino}
\affiliation{IBM Quantum, IBM Research – Zurich, 8803 Rüschlikon, Switzerland}
\author{Giuseppe Murante}
\affiliation{INAF - Osservatorio Astronomico di Trieste, via G.B. Tiepolo 11, 34143 Trieste, Italy\\
ICSC - Italian Research Center on High Performance Computing, Big Data and Quantum Computing\\}
\author{Stefano Borgani}
\affiliation{Dipartimento di Fisica dell'Universit\`a di Trieste, via Tiepolo 11, I-34131 Trieste, Italy\\
INAF - Osservatorio Astronomico di Trieste, Trieste, Italy\\
IFPU, Institute for Fundamental Physics of the Universe, Trieste, Italy\\
ICSC - Italian Research Center on High Performance Computing, Big Data and Quantum Computing\\}
\author{Ivano Tavernelli}
\email{ita@zurich.ibm.com}
\affiliation{IBM Quantum, IBM Research – Zurich, 8803 Rüschlikon, Switzerland\\}
\vspace{0.5truecm}
\date{\today}

\begin{abstract}
Cosmological simulations describing the evolution of density perturbations of a self-gravitating collisionless Dark Matter (DM) fluid in an expanding background, provide a powerful tool to follow the formation of cosmic structures over wide dynamic ranges. The most widely adopted approach, based on the N-body discretization of the collisionless Vlasov-Poisson (VP) equations, is hampered by an unfavourable scaling when simulating the wide range of scales needed to cover at the same time the formation of single galaxies and of the largest cosmic structures. On the other hand,
the dynamics described by the VP equations is limited by the rapid 
increase of the number of resolution elements (grid points and/or particles) which is required to simulate an ever growing range of scales. 
Recent studies showed an interesting mapping of the $6$-dimensional$+ 1$ ($6D+1$) VP problem into a more amenable  $3D+1$ non-linear Schr\"odinger-Poisson (SP) problem for simulating the evolution of DM perturbations. 
This opens up the possibility of improving the scaling of time propagation simulations using quantum computing. 
In this paper, we introduce a quantum algorithm for simulating the Schrödinger-Poisson (SP) equation by adapting a variational real-time evolution approach to a self-consistent, non-linear, problem. To achieve this, we designed a novel set of quantum circuits that establish connections between the solution of the original Poisson equation and the solution of the corresponding time-dependent Schr\"odinger equation. We also analyzed how nonlinearity impacts the variance of observables. Furthermore, we explored how the spatial resolution behaves as the SP dynamics approaches the classical limit ($\hbar / m \rightarrow 0$) and discovered an empirical logarithmic relationship between the required number of qubits and the scale of the SP equation ($\hbar/m$). This entire approach holds the potential to serve as an efficient alternative for solving the Vlasov-Poisson (VP) equation by means of classical algorithms.
\end{abstract}

\maketitle

~
%\newpage
%~
%\newpage
%~

\section{Introduction}
A number of astrophysical and cosmological observations consistently point toward the definition of the so-called standard cosmological model~\cite{Planck2020}. In this model, the mass-energy content of the Universe is made by about 70\% of an unknown form of Dark Energy (DE), which accounts for the accelerated cosmic expansion, by about 25\% of an unknown form of collisionless non-baryonic Dark Matter (DM), while only the remaining $\sim 5\%$ is made of ordinary baryonic matter. In addition, viable models of galaxy formation require DM particles to be cold (CDM), i.e. with negligible streaming velocities. With the further observational evidences for DE to be consistent with a cosmological constant term ($\Lambda$) in the Einstein field equations, all this leads to the definition of the standard $\Lambda$CDM cosmological model \cite{Weinberg2013}. While the exact nature of cosmic dark constituents remains so far elusive, it is widely accepted that the gravitational instability of the tiny CDM density perturbations imprinted in the primordial Universe drive the formation of cosmic structures, from kiloparsec (kpc) scales relevant for galaxy formation, to the Gigaparsec (Gpc) scales of the global cosmic web \cite{MvdBW}. Describing in detail the evolution of such DM perturbations within a DE-dominated expanding background, and comparing the predictions to observational data is crucial to shed light on the nature of DM and DE. The most widely adopted approach to address the study of the gravitational instability of density perturbations in a collisionless fluid is by adopting the N-body discretization of the evolution of fluid phase-space distribution function described by the Vlasov-Poisson (VP) system of equations \cite{Springel2016}.

In its most straightforward implementation, the N-body method explicitly computes the gravitational interaction between each pair of the N particles, which discretize the fluid, thus implying a $N^2$ scaling with the number of resolution elements. While different methods, based on different levels of numerical approximation, have been introduced to speed-up these computations, still they are currently hampered by the unfavorable scaling of the available classical algorithms with the respect to system sizes. Furthermore, we should keep in mind that the N-body discretization of the phase-space structure of the fluid is also an approximation to reduce the dimensionality of the problem to a treatable level.

A recent work by Mocz et \textit{al.}~\cite{mocz_schrodinger-poissonvlasov-poisson_2018} showing numerical correspondence between the $6D + 1$ Vlasov-Poisson (VP) and the $3D+1$ Schr\"ondiger-Poisson (SP) equations for cosmological simulation revived the interest in simulating and studying various form of dark matter, which can be modelled by the SP equation \cite{schwabe2020simulating,mocz_towards_2021,angulo2022large}. 
In fact, the SP equation has also a direct physical interpretation of the  \textit{so-called} axion model, which postulates the presence of scalar particles as constituents of dark matter.
In the ultra-light particle mass limit, this model is known as fuzzy dark matter (FDM)~\cite{hu2000fuzzy}. 
This correspondence opens up the possibility of using quantum algorithms (QA) for the investigation of dark matter dynamics, as it was already demonstrating that QA can reduce the scaling complexity for the solution of quantum mechanical problems in many-body physics and quantum chemistry~\cite{feynman_simulating_1982,tacchino2020quantum,Miessen2023}. 

%--------------------------------
\begin{figure}[H]
    \centering
    \includegraphics[width = \columnwidth]{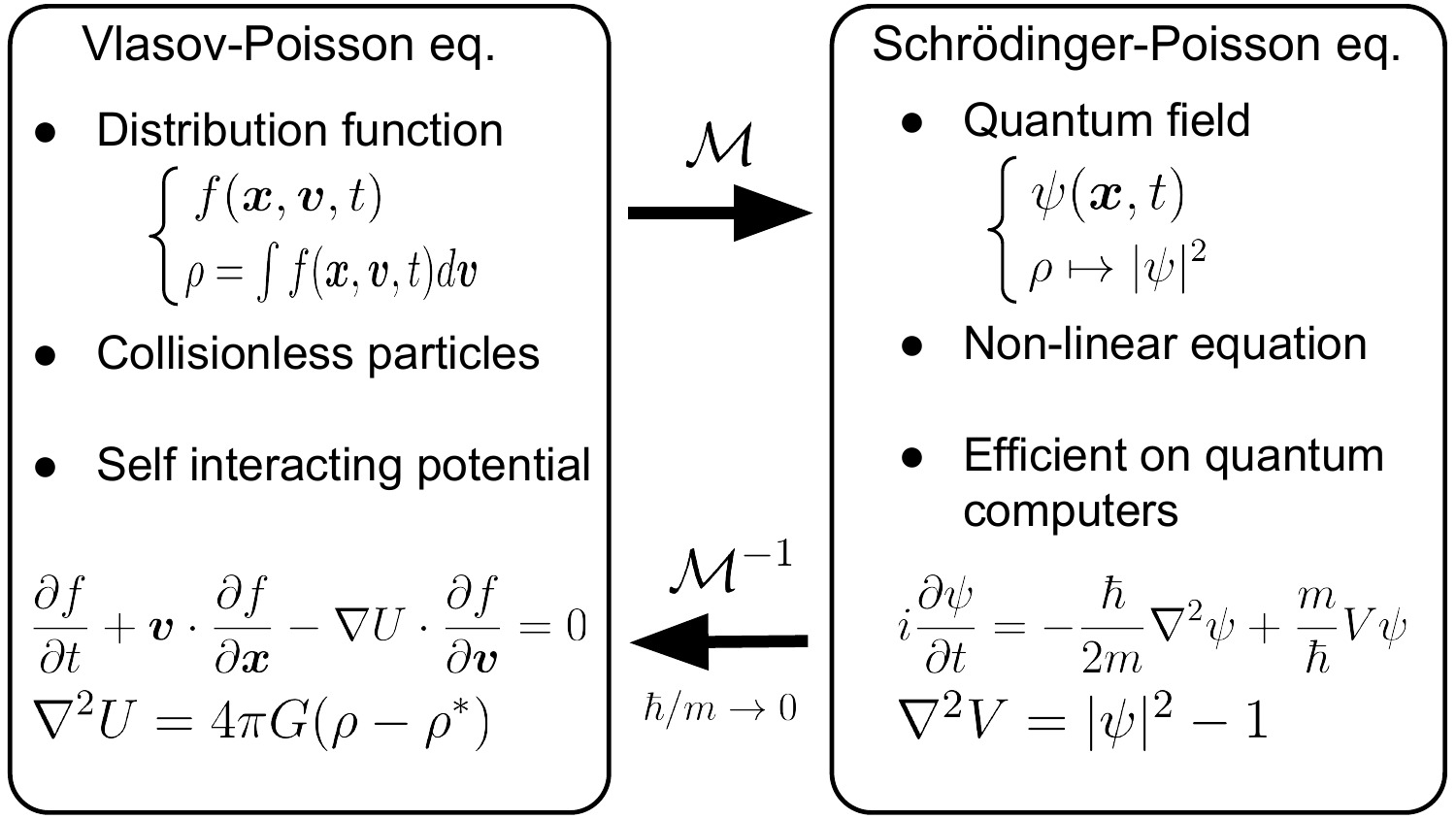}
    \caption{ \small 
     Mapping, $\mathcal{M}$, of the classical $N-$body Vlasov problem into the corresponding quantum Schr\"odinger Poisson formulation obtained through nonlocal manipulation (e.g, Husimi smoothing~\cite{Husimi}). 
     Detail on the mapping $\mathcal{M}$ and its inverse are given in~\cite{mocz_schrodinger-poissonvlasov-poisson_2018, 1993kaiser}. 
    }
    \label{fig:mapping}
\end{figure}

%-------------------------------------------------------------

More generally, we propose a scalable quantum algorithm for the simulation of the time propagation of non-linear Schr\"odinger-like equations of the form
\begin{equation}\label{eq:Sch_equation}
    i \frac{\partial}{\partial t} \Psi  = 
    H[\Psi] \, \Psi
\end{equation}
where $H[\Psi]$ indicates the functional dependence of the Hamiltonian from the system wavefunction. 

In this work, we explore the challenges arising in the implementation of cosmological simulations on quantum devices.
The dynamics is governed by the SP equation, where a self-gravitating potential introduces nonlinearities in the problem. 
The mapping of the nonlinear problem onto a quantum device is solved using a classical-hybrid variational algorithm similar to the one proposed in Lubasch et al.~\cite{lubasch_variational_2020}.
The evolution of the wavefunction is carried out using a variational time evolution (VTE) approach, tailored for nonlinear self-consistent problems defined on a grid, which allows for an exponential saving in computational memory resources through the encoding of $N$ grid points in $\log_2(N)$ qubits. 
Building on~\cite{ollitrault_quantum_2022}, we adapt the VTE algorithm to the case where the potential is given by a variational ansatz, proposing quantum circuits for the evaluation of the required matrix elements whose depth scaling is polynomial with the number of qubits and the number of sample required for a desired accuracy scales polynomially with the number of grid points $N$.

We investigate the behavior of spatial resolution as the SP dynamics converges towards the classical limit ($\hbar / m \rightarrow 0$). Our investigation unveiled an empirical logarithmic correlation between the required number of qubits and the scale of the SP equation ($\hbar/m$). 

This work is structured as follows. 
In Section \ref{sec:Theory and methods} we describe the mapping of the cosmological SP equation on a quantum computer, including a discussion of the strategies that must be adopted in the latter for the description of non-linear problems.

Section~\ref{sec:the_algorithm} is devoted to the description of the VTE algorithm for self-consistent nonlinear problems, including a discussion on the quantum circuit implementation. 
Numerical simulations for a one dimensional 5-qubit (i.e., 32 grid points) system will be given in Section~\ref{sec:results}. 
The results include an analysis of the time-evolution obtained with different choices of physical parameters interpolating between the pure quantum regime and a classical, $\hbar/m \rightarrow 0$, limit.
A study on the resolution convergence in this classical regime is also presented.
Finally, we discuss the computational costs of our quantum algorithm 
and the conditions for potential quantum advantage. We draw our main conclusions in Section~\ref{conclusions}

\section{Theory and Methods}\label{sec:Theory and methods}
\subsection{History of the Schr\"odinger-Poisson equation}\label{subsec:cosmological intro}

Under the fluid assumption, the phase-space distribution of
  massive CDM particles at time $t$ is described by the
  \textit{distribution function} $f(\mathbf{x}, \mathbf{v}, t)$, where
  $\mathbf{x}, \mathbf{v} \in \mathrm{R}^3$ are the positions and
  velocities of the particles, so that $f\,d\mathbf{x}\,d\mathbf{v}$
  describe the phase-space density within the 6D volume element
  $d\mathbf{x}\,d\mathbf{v}$, so that the density field in
  configuration space is given by
  $\rho(\mathbf{x},t)=\int f(\mathbf{x}, \mathbf{v}, t)\,d\mathbf{v}$. Under the assumption of a
  collisionless fluid, the evolution of the distribution function
  obeys a continuity equation in phase-space,
  $df(\mathbf{x}, \mathbf{v}, t)/dt=0$. If the fluid is
  self-gravitating, then the Poisson equation,
  $\nabla^2U(\mathbf{x},t)=4\pi G \rho(\mathbf{x},t)$ (with $G$ the
  Newton's gravitational constant) provides the
  relationship between the density field and the gravitational
  potential $U$ \cite{Binney}.  Simulations of cosmic structure
  formation within a $\Lambda$CDM model aim at solving this
  \textit{Vlasov-Poisson} system of equations, once initial conditions
  on position and velocity of the particles
  $f(\mathbf{x}, \mathbf{v}, t_0)$ are assigned to represent an
  ensemble realization of a given cosmological model
  \cite{DolagBorgani2008}. As such, the VP equations must be solved in
  $6D+1$ dimensions. The high dimensionality of this problem makes it
  very hard to tackle when a high spatial resolution is needed, as
  usual in modern cosmological simulations.
  \\
  A widely used approach to reduce the dimensionality of the problem
  is to model the initial DM distribution as an ensemble of
  collisionless massive particles interacting only through
  self-gravity. Such a set of particles formally obeys to the {\it
    Euler-Poisson} (EP) equations, a closure of the VP equations
  obtained by asking that the distribution function is single-valued
  in space.  Classically the evolution is carried out using N-body
  \cite{many_body1, many_body2, many_body3_review} or fluid approaches
  \cite{fluid_approach}.
  \\
  The $N$-body approach \cite{Garny_2018, many_body1, many_body2} best
  approximates the analytic solution of the system (each DM particle
  has a single-valued velocity; at large scales however they can
  cross, as the VP equations require) and usually presents no
  singularities. However, it requires much more computational
  resources than the fluid one.  On the other hand, the fluid method,
  that directly solves the EP equations, manages to reduce the
  dimensionality of the problem from $6D+1$ to $3D+1$, but presents
  singularities
  \cite{mocz_schrodinger-poissonvlasov-poisson_2018, fluid_approach}.
\\
  The potential limitations of both the N-body and the fluid methods
  clearly demonstrates that finding an alternative and efficient way
  to solve the VP equations would provide a significant conceptual and
  computational benefit for the numerical study of cosmic structure
  formation.
\\
Within this context, the Schr\"odinger-Poisson (SP) equations,
i.e. the coupling of the Schr\"odinger equation with a
self-interacting potential obeying the Poisson equation, have recently
been proved to recover in the classical limit $\hbar/m \to 0$ the
dynamics of the VP equations
\cite{mocz_schrodinger-poissonvlasov-poisson_2018, LANGE1995331,
  davies_test-bed_1996}.  Such an approach was first introduced in
ref. \cite{ruffini_systems_1969} as the non-relativistic limit of the
Einstein field equations with a scalar boson field as source.
\\
%%%%%%%%%%%%%%%%%%%%%%%%%%%%%%%%%%%%%
The procedure known as the \textit{Schr\"odinger method} (SM) maps the initial classical distribution $f(\mathbf{x}, \mathbf{v}, t_0)$ to the quantum wavefunction
    $\Psi(\mathbf{x}, t_0)$ by means of a nonlocal operation. 
    Details about this procedure are given in \cite{mocz_schrodinger-poissonvlasov-poisson_2018}; here we just provide a brief overview of the method.
    We consider two primary scenarios. In instances where the initial distribution function is characterized by a cold or single-valued stream, meaning that a unique velocity corresponds to each point, it is possible to directly reconstruct the phase $S$ of the quantum wavefunction $\Psi = \sqrt{\rho} \, \exp(iS/\hbar)$, where $\rho = \int d\boldsymbol{v} f(\boldsymbol{x}, \boldsymbol{v},t)$ through the solution of the Poisson problem,
    \begin{equation}
        \label{eq: phase for wavefunction}
        \boldsymbol{\nabla} \cdot \boldsymbol{v} = \nabla^2 S / m \, ,
    \end{equation}
    where the scale $\hbar/m$ emerges as an effect of the quantization. 

    In the scenario involving multi-streams or warm initial conditions, where a single grid point may correspond to two or more different velocity values, the situation becomes more complex. This complexity arises because the densities do not precisely coincide, and the quantum wavefunction incorporates interference patterns. In this case, the mapping from phase-space distribution function to wavefunction reads
    \begin{equation}
        \label{eq: wavefunnc from warm distribution}
        \Psi(x) \propto \sum_{\boldsymbol{v}} \sqrt{f(\boldsymbol{x}, \boldsymbol{v}}) e^{\frac{i}{\hbar}  m \, \boldsymbol{x} \cdot \boldsymbol{v}  + 2\pi i \phi_{rand , \boldsymbol{v}}} \Delta\boldsymbol{v} \, ,
    \end{equation}
    where, we sum over sampled velocities $\boldsymbol{v}$, each one with an associated random phase $\phi_{rand, \boldsymbol{v}} \in [0,2\pi)$ to ensure uncorrelated phases for each fluid velocity.

%%%%%%%%%%%%%%%%%%%%%%%%%%%%%%%%%%%%%%%
%through a Husimi transformation \cite{Husimi} with an accuracy of $O(\hbar / m), O((\hbar / m)^2)$.  
The wavefunction then evolves according to the SP system of equations
\begin{align}
\label{eq:fuzzy_SP}
        &i \hbar \frac{\partial \Psi}{\partial t}=-\frac{\hbar^{2}}{2 m} \nabla^{2} \Psi+m U \Psi\,;
        \\
        &\nabla^{2} U = 4 \pi G(\rho - \rho^*)\,.
\end{align}
Here we have chosen to use a density contrast $\rho - \rho^*$ as source of the gravitational potential,
where $\rho^*$ represents the average density over the volume considered.
We note that in this approach Eq.~\eqref{eq:fuzzy_SP} describes a density field, not a particle's wavefunction. Note also that the constant $\hbar$ and $m$ are not the Planck constant and the mass of the particle but are related respectively to the quantum and classical effects (see discussion below in Sect. \ref{subsec:problems_nonlinear} and 
%Supplemental Material 
in the Appendix~\ref{supp: SP rescaling}
for details about the scale of the equation).
%%%%%%%%%%%%%%%%%%%%%%%%%%%%%%%%%%%%%%%%%%

    Once the Schr\"odinger-Poisson evolution is completed, the distribution function can be extracted from the final wavefunction using the Husimi procedure, which is a smoothed version of the Wigner quasi-probability distribution.  A similar approach can be found in \cite{Vlasov_bertrand} for the solution of the Vlasov equation with electromagnetic fields in plasma physics applications. In a $3D$ context, this operation can be seen as the spatial smoothing of the wavefunction $\Psi$ with a Gaussian filter of width $\eta$. Additionally, it involves a Fourier-like transformation to extract momentum information,
    \begin{align}
        \label{eq: Husimi smoothing}
        &\tilde{\Psi}(\boldsymbol{x}, \boldsymbol{p},t; \eta)
        = \left( \frac{1}{2\pi \hbar} \right)^{3/2} \left( \frac{1}{\pi \eta^2} \right)^{3/4}
        \\ & 
        \nonumber
        \times \int d^{3}r \, \Psi(\boldsymbol{r}, t) \exp{\left( -\frac{(\boldsymbol{x} - \boldsymbol{r})^2}{2\eta^2} - i \frac{\boldsymbol{p} \cdot (\boldsymbol{r} - \boldsymbol{x}/2)}{\hbar}\right)} \,.
    \end{align}
    The squared module of the wavefunction in Eq.~\eqref{eq: Husimi smoothing} yields a result that closely approximates the desired distribution function~\cite{mocz_schrodinger-poissonvlasov-poisson_2018, Husimi}.

%%%%%%%%%%%%%%%%%%%%%%%%%%%%%%%%%%%%%%%%%%
As a side note, we remind that the SP equations has been already
used in the numerical study of cosmic structure formation to study the
dynamics of the Fuzzy Dark Matter (FDM) perturbations
\cite{BaldiNori2018,Mocz2019}. This class of DM candidates emerges as
the ultra-light mass limit of a scalar bosonic field, whose particles
are known as axions. In this case $\hbar$ represents in fact the
actual Planck constant and $m$ the mass of the axion-like
particles. The characteristic scale of the problem is the ratio
$\hbar/m$: at smaller scales the dynamics is influenced by quantum
effects as quantum pressure, while at larger scales, this effect
becomes negligible and the classical Cold Dark Matter (CDM) limit is
recovered.
%%%%%%%%%%%%%%%%%%%%%%%%%%%%%%%%%%%%%%

\subsection{The nonlinear SP equation on quantum computers}\label{subsec:problems_nonlinear}
We consider a complex wavefunction
$\Psi(\mathbf{x}, t)$ (with $\mathbf{x} \in \mathbb{R}^3$) defined in such a way that
$|\Psi|^2 = \rho / \rho^*$. The following normalization emerges naturally from the definition of the volume-mean density $\rho^*$ 

%normalized over the volume $\mathcal{V}$ as 
\begin{equation}
\label{eq: normalization volume}
\frac{1}{\mathcal{V}} \int d\mathcal{V} \, |\Psi|^2 = 1,
\end{equation}
The SP equation of interest (see diagram in Fig.~\ref{fig:mapping}) 
assumes the general form  
\begin{equation}\label{eq:Sch_equation_our}
    i \frac{\partial}{\partial t} \Psi(\mathbf{x}, t)  = 
    \left(
    -\frac{\lambda}{2}\nabla^{2}  +
    \frac{1}{\lambda} %V(\mathbf{x}, t) 
    V[\Psi(\mathbf{x}, t)]
    \right)
    \Psi(\mathbf{x}, t) \, .
\end{equation}
with the self-interacting potential 
$V[\Psi]$
defined as 
\begin{equation}
    \label{eq:poisson_equation}
    \nabla^{2} V[\Psi]= \nabla^{2} V(\mathbf{x}, t) = | \Psi(\mathbf{x}, t) |^{2}-1 \,. 
\end{equation}

Here $\lambda = \hbar/m$ is the intrinsic scale of the problem 
%\cite{mocz_schrodinger-poissonvlasov-poisson_2018} 
and $V[\Psi]$ is a redefinition of the self interacting potential $U[\Psi]$ that renders the Poisson equation dimensionless. 
We use square brackets, e.g., $V[\Psi]$, to denote functional dependence.
Details on how to recover Eqs.~\eqref{eq:Sch_equation_our},~\eqref{eq:poisson_equation} from Eq.~\eqref{eq:fuzzy_SP} are given in the 
Appendix~\ref{supp: SP rescaling}. 
%Supplemental Material at [URL will be inserted by publisher].
%
This set of equations can be seen as a time-dependant Schr\"odinger-like equation (TDSE), where the self-interacting nature of the potential in Eq. (\ref{eq:poisson_equation}) causes the dynamics of the system to be strongly nonlinear. 
It features two main processes, whose intensity are regulated by 
the magnitude of $\lambda$. 
We observe that if $\lambda\to \infty$ the potential term vanishes, 
leaving only the free Schr\"odinger equation which leads to diffusion~\cite{2021schrodinger} (however, due to the imaginary coefficient
$i\lambda/2$, the Schr\"odinger equation cannot be strictly classified as a diffusion equation). 
In this case we expect to see a spatial smoothing of the density distribution. 
In the opposite limit, when $\lambda\to 0$, the potential term dominates: this should cause the collapse of the distribution followed by a series of peaks and caustics. As such, this can be seen as the onset of the classical regime of gravitational instability \cite{mocz_schrodinger-poissonvlasov-poisson_2018}.

While quantum computation proved to be efficient in solving linear partial differential equations (PDEs)~\cite{kassal_polynomial-time_2008, quantum_linear_eqution, aspuru-guzik_simulated_2005} problems arise when dealing with \textit{nonlinear} equations  due to the intrinsic \textit{linearity} of the quantum 
computation formalism \cite{feynman_simulating_1982,barenco_elementary_1995, nielsen_chuang_2010}.
Two main challenges are associated to the non-linearity of Eq.~\eqref{eq:Sch_equation_our}. 
The first one is related to the fact that quantum states are usually prepared and evolved through unitary operations. This preserves the well-known probability-like normalization of the quantum register: 
$\langle \psi\ket{\psi} = 1$.
Thus, the \textit{physical} wavefunction $\ket{\Psi}$, that solves Eq.~\eqref{eq:Sch_equation_our}, and the generic \textit{quantum} state on the quantum register $\ket{\psi}$ live in two different Hilbert spaces. We will give more details on this subject in  Section~\ref{subsec_quantum-approach-to-sp}. The second complication is related to the self-consistency of the problem, 
which forces us to look at alternative time evolution algorithms than Trotter-based expansions~\cite{somma_quantum_2016, ollitrault_non-adiabatic_2020}. 

To address both issues, in this work we propose a variational time evolution algorithm specifically adapted to the nonlinearity of the problem, which relies on the development and the application of novel quantum circuits described in Sec.~\ref{sec:the_algorithm}.
%%%%%%%%%%%%%%%%%%%%%%%%%
\begin{figure}
    \resizebox{\columnwidth}{!}
    {
     \begin{quantikz}
         \lstick{$\ket{0}$} &\gate{R_y(\theta_0)}
         \gategroup[wires=3,steps=1,style={inner sep=6pt}]{rotation}
         &[4mm]\ctrl{1}
         \gategroup[wires=3,steps=2,style={inner sep=6pt}]{entangling}
         &\qw &[2mm]\qw &\gate{R_y(\theta_3)}
         &[4mm]\ctrl{1} 
         &\qw &[2mm]\qw &\gate{R_y(\theta_6)}
         \\
         \lstick{$\ket{0}$} &\gate{R_y(\theta_1)} &\targ{} &\ctrl{1} &\qw &\gate{R_y(\theta_4)}
         &\targ{} &\ctrl{1} &\qw &\gate{R_y(\theta_7)}
         \\
         \lstick{$\ket{0}$} &\gate{R_y(\theta_2)} &\qw &\targ{} &\qw  &\gate{R_y(\theta_5)}
         &\qw &\targ{} &\qw  &\gate{R_y(\theta_8)}
    \end{quantikz}
    }         
    \caption{ Example of a 3-qubit $\mathrm{R}_y$-$\mathrm{CNOT}$ ansatz circuit used for the  wavefunction 
    $\ket{\Phi_{\tilde{V}}}$ used to evaluate the potential according to Eq.~\eqref{e1: potential normalization}. This circuit has 3 rotational blocks $U_{rot}$ and 2 entangling blocks $U_{ent}$ with \textit{linear} entanglement. The output function is parameterized through the nine real parameters $\boldsymbol{\theta}$ such that $U(\boldsymbol{\theta})\ket{\mathbf{0}} = |\Phi_{\Tilde{V}}(\boldsymbol{\theta}) \rangle$; in this case the number of parameters exceeds the Hilbert space dimension $2^3 = 8$.
    }
    \label{fig:ry_cnot}
\end{figure}
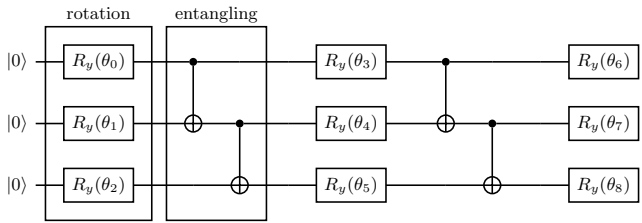

\subsection{The quantum computing approach to the SP equation}
\label{subsec_quantum-approach-to-sp}
A first attempt to solve the nonlinear SP equation was given by Mocz and Szasz~\cite{mocz_towards_2021}. Such a solution is fully variational and makes use of a finite difference optimization of the potential and of the system wavefunction evaluated at two subsequent time steps. 
The variational nature of this approach also allows one to bypass the costly solution of the  Poisson equation in Fourier space in favour of a variational optimization of the potential as implemented in a separate qubit register.

In this work, we propose a novel set of quantum circuits that enable the implementation of a different strategy based on a adapted variational time-dependent quantum algorithm for the propagation of the variational parameters defining the system wavefunction (See Section~\ref{subsubsec_wavefunction}). 
This enables a more rigorous implementation of the wavefunction dynamics, avoiding the instabilities implicit in most VQE optimization procedures (e.g., slow convergence due to the trapping in local minima and barren plateaus~\cite{Lorenz2021causalcompositional, gacon2023variational}). On the other hand the VTE algorithm comes at the cost of evaluating additional matrix elements for the solution of the equation of motion for the wavefunction parameters. 
%%%%%%%%%%%%%%%%%%%%%%%%%%%%%%%%%%
\subsubsection{Grid-based representation of the system wavefunction}\label{subsubsec_wavefunction}
A typical space discretization associated to problems in first quantization  \cite{somma_quantum_2016, lubasch_variational_2020, mocz_towards_2021, ollitrault_non-adiabatic_2020, ollitrault_quantum_2022}  approximates a continuous space with a grid.
In 1D, a line of length $L$ is divided in arbitrary $N$ equidistant points. 
%, which can be encoded in $n=\lceil \log(N) \rceil$ qubits. 
For each grid point $x_j$ we have $\Psi_j \simeq \Psi(x_j)$, with $j \in \{0, 1,..., N-1\}$ and periodic boundary conditions $\Psi_N = \Psi_0$.
\\
With a $n$-qubit quantum register, one can generate a quantum state $|\psi\rangle$ belonging to a $N$-dimensional Hilbert space, where $N = 2^n$. Making use of such logarithmic encoding, only $n = \log_2N$ qubits are needed to describe a $N$-point grid. A generic state $|\psi\rangle$ can hence be represented on a quantum register as a superposition of computational basis states,
\begin{equation}\label{eq: generic_wavefunc}
    |\psi\rangle = \sum_{k=0}^{N-1} \psi_j |\mathrm{bin}(j)\rangle,
\end{equation}
where $\mathrm{bin}(j)$ is the binary representation of the grid position $j$ and $\psi_j \in \mathbb{C}$ is the associated amplitude or weight, such that the probability distribution of measuring the different basis states (i.e., different positions on the grid) is normalized as
$
    \langle \psi| \psi \rangle = \sum_{j = 0}^{N-1} |\psi_j|^2 = 1.
$
By combining this relation with the discretization of Eq.~\eqref{eq: normalization volume}, we can establish a correspondence between the approximated physical wavefunction on the grid point $x_j$ and the corresponding coefficient of the $j$-th basis $\ket{\mathrm{bin}(j)}$ in Eq.~\eqref{eq: generic_wavefunc}, such that $\Psi_j = \sqrt{N} \psi_j$.

The dynamics of the system wavefunction is  described by means of a time-dependent variational approach~\cite{yuan_theory_2019}. 
To this end, we define a quantum trial state $\ket{\psi(\boldsymbol{\theta}(t))}$, parameterized by a set of (time-dependent) variables $\boldsymbol{\theta}(t) = \{ \theta_1(t),...,\theta_{M_p}(t) \}$, which evolve according to well-defined equations of motion~\cite{yuan_theory_2019}. The initial state is prepared through a suitable choice of a  parameterized unitary (quantum circuit)  $U(\boldsymbol{\theta}(0))$. 
An explicit circuit example is shown in %Fig.~\ref{fig:ry_cnot}.
Fig.~\ref{fig: derivatives circuits}.
Using the previous relation between $\Psi_j$ and $\psi_j$, we can describe the time evolution of the physical state 
\begin{equation}
\label{eq:physical_quantum_relation}
    \ket{\Psi(\boldsymbol{\theta}(t))} = \sqrt{N} \ket{\psi(\boldsymbol{\theta}(t))}.
\end{equation}
using the updated parameters $\boldsymbol{\theta}(t)$ (see~\ref{subsubsec:time_evolution}).
%
%%%%%%%%%%%%%%%%%%%%%%%%%%%%%%%%%%
%%%%%%%%%%%%%%%%
\subsubsection{Variational time propagation with nonlinearities}\label{subsubsec:time_evolution}   
The trial wavefunction $ \ket{ \psi(\boldsymbol{\theta}(t)) }$ is evolved adapting the VTE algorithm proposed in Ref.~\cite{ollitrault_quantum_2022}  to the case where the potential is self-consistent with the wavefunction and needs to be re-evaluated at each timestep. In VTE, the dynamics is tracked on the manifold spanned by the time-dependent parameters $\boldsymbol{\theta}(t)$ used to describe the trial wavefunction.

For a system evolving under the action of a Hamiltonian $\mathcal{H}$, we derive, from the McLachlan variational principle~\cite{yuan_theory_2019,ollitrault_quantum_2022}, a set of equations of motion (EOM) of the form
\begin{equation}
    \label{eq: McLach eq. of motion}
    \mathrm{M} \, \dot{\boldsymbol{\theta}}=\mathrm{B},
\end{equation}
where 
\begin{align}
    \label{eq: McLach M}
    \mathrm{M}_{ kl}&=\Re\left\{ 
    \left\langle\partial_{\theta_{k}} \Psi \mid \partial_{\theta_{l}} \Psi\right\rangle
    -
    \left\langle\partial_{\theta_{k}} \Psi \mid \Psi\right\rangle\left\langle\Psi \mid \partial_{\theta_{l}} \Psi\right\rangle
    \right\} \\
    \label{eq: McLach V}
    \mathrm{B}_k&=\Im\left\{\left\langle\partial_{\theta_{k}} \Psi|\mathcal{H}| \Psi\right\rangle-\left\langle\partial_{\theta_{k}} \Psi \mid \Psi\right\rangle\langle\Psi|\mathcal{H}| \Psi\rangle\right\}
\end{align}
%%%%%
%--------------------
\begin{algorithm}[H]
    \caption{alg: VTE for Self-consistent potential.}\label{alg: VTE scheme}
        \begin{algorithmic}
            \State $i=0$ \Comment{Initialization}
            \State $\psi_{t_0} \gets \boldsymbol{\theta}_{t_0} $
            \While{$\{\boldsymbol{\phi}_{t_0}\}$ not converged}
                \State $ V_{t_0} \gets \boldsymbol{\phi}_{t_0} $   
                \State cost = $||V_{t_0} - |\Psi_{t_0}|^2 + 1 ||^2$
            \EndWhile
            \For{$i = 0,..., N_t - 1$}:
                \State evaluate $\mathrm{M}_{k, l}(\boldsymbol{\theta}_i), \mathrm{B}_k(\boldsymbol{\theta}_i, \boldsymbol{\phi}_{t_i})$ \Comment{V.T.E}
                \State $\boldsymbol{\theta}_{t_{i+1}}  \gets
                \mathrm{M} \cdot \dot{\boldsymbol{\theta}}_{t_i} = \mathrm{B}$
                \State $\psi_{t_{i+1}} \gets \boldsymbol{\theta}_{t_{i+1}}$
                \\
                \While{$\{\boldsymbol{\phi}_{t_{i+1}}\}$ not converged}     \Comment{Pot. Opt.}
                    \State $ V_{t_{i+1}} \gets \boldsymbol{\phi}_{t_{i+1}} $   
                    \State cost = $||V_{t_{i+1}} - |\Psi_{t_{i+1}}|^2 + 1 ||^2$
                \EndWhile
            \EndFor
        \end{algorithmic}
\end{algorithm}
%--------------------
%
with 
\begin{equation}
\label{eq:Hamiltonian}
\mathcal{H}[\Psi]=\left(
    -\frac{\lambda}{2}\nabla^{2}  +
    \frac{1}{\lambda} %V(\mathbf{x}, t) 
    V[\Psi]
    \right)
\end{equation}
as defined in Eq.~\eqref{eq:Sch_equation_our}. 
The dependence of $\Psi$ on the parameters $\boldsymbol{\theta}(t)$ is implicit. 
Note that to capture the exact evolution comprehensive of nonlinear effects, the terms in Eqs.~\eqref{eq: McLach M} and \eqref{eq: McLach V} are rescaled according to Eq.~\eqref{eq:physical_quantum_relation}.
The main obstacle to the application of such method is the evaluation of the term 
    $\Im\left\langle\partial_{\theta_{k}} \Psi|\mathcal{H}| \Psi\right\rangle$ in Eq.~\eqref{eq: McLach V}. The difficulty stands in the fact that the self-consistent potential does not have a standard form, but it depends on the system wavefunction. The evaluation of this term is made possible by application of the novel quantum circuit scheme discussed in Sec.~\ref{sec:the_algorithm}.

%%%%%%%%%%%%%%%%%%%%%%%%%%%%%%%%%%%%%%%%%%%%%%%%%%%
%
\subsubsection{Optimization of the potential} 
\label{subsubsec_potential_optimization}
As anticipated in Sec.~\ref{subsec:problems_nonlinear}, the functional dependence of the potential on the system wavefunction, $\Psi$, brings a further level of complexity into the dynamics of the system.
While classically, the solution of the Poisson equation (\ref{eq:poisson_equation}) for a generic wavefunction $\Psi$ can easily be found using a spectral method in Fourier space~\cite{mocz_towards_2021}, such strategy is not practical on near-term quantum computers, as it would require rather deep circuits~\bibnote{One way of  implementing a quantum spectral method for the solution of the SP equation would require a quantum Fourier transform ($QFT$) to move in the momentum space. Then a circuit able to reproduce $|\psi|^2-1$ would need to be followed by one able to divide for the squared momenta $k^2$. In the end a $QFT^{-1}$ would return the exact potential on the quantum register.}.
We instead resort to a variational approach.
Hence, we introduce a second set of parameters 
$\boldsymbol{\phi}(t) = \{ \phi_V(t), \tilde{\phi}_1(t),..., \tilde{\phi}_L(t) \}$ 
describing a quantum state 
%%%
\begin{equation}
\label{eq: potential structure state}
    \ket{\Phi_V(\boldsymbol{\phi})} = \phi_V \ket{\Phi_{\Tilde{V}} (\tilde{\boldsymbol{\phi})}},
\end{equation}
such that the potential can be obtained as
\begin{equation}
\label{eq: potential structure}
%    \ket{\Phi_V(\boldsymbol{\phi})} = \sum_{j=0}^{N-1}  V_j(\phi(t)) \, \ket{j} \,
%    = \phi_V \sum_{j=0}^{N-1}  \tilde{V}_j(\phi(t)) \, \ket{j} \, ,
        \ket{\Phi_V(\boldsymbol{\phi})} = \sum_{j=0}^{N-1}  V_j(\phi(t))  \ket{j} 
    = \phi_V \sum_{j=0}^{N-1}  \tilde{V}_j(\phi(t)) \ket{j} \, ,
\end{equation}
where the index $j$ in $V_j(\phi(t))$ labels the grid position $\mathbf{x}_j$ associated to the bit string $\text{bin}(j)$.
In Eq.~\eqref{eq: potential structure}  the parameter $\phi_V$ \cite{mocz_towards_2021} ensures the normalization of the potential,
\begin{equation} \label{e1: potential normalization}
\braket{\Phi_{\tilde{V}}(\tilde{\phi}) | \Phi_{\tilde{V}}(\tilde{\phi})} = 
\sum_{j=0}^{N-1} |\tilde{V}_j(\phi(t))|^2 = 1 \,, \, \forall t \, .
\end{equation}
The potential can therefore be interpreted as a functional of the circuit parameters, 
$V_j(\boldsymbol{\phi}(t))$. 
The parameters are iteratively updated to minimize the distance between the parameterized potential and the one arising from Eq.~\eqref{eq:poisson_equation}:
\begin{equation}
\label{eq:analytic_cost_potential}
    min_{\boldsymbol{\phi}}
    \left(
    \sum_{j=0}^{N-1} 
    \left(
    {\nabla^2V_j(\boldsymbol{\phi}) - |\Psi_j(\boldsymbol{\theta})|^2 + 1}
    \right)^2 
    %\left(\sum_k {\nabla^2V_k(\boldsymbol{\phi}) - |\Phi_k(\boldsymbol{\theta})|^2 + 1}\right)
    \right) \, .
\end{equation}
Details about the terms appearing in Eq.~\eqref{eq:analytic_cost_potential} are given in Appendix~\ref{sec: potential}.
When the optimization converges, the function $V_j(\boldsymbol{\phi}(t))$ approximates the exact potential $V(\mathbf{x}, t)$ with $\mathbf{x} \in \{ \mathbf{x}_j \}$ corresponding to the parameterized wavefunction $\Psi(\boldsymbol{\theta}(t))$ at a specific time $t$. 
%
%*************************************
%%%%%%%%%%%%%%%%%%%%%%%%%%%%%%%%%%%%%%%%%%%%%%%%
%       %%%%%%%%%%%%%%%%%%%%%%%%%%%%%%%%%%%%%%%%%%%%%%%%%%%%%
%%%%%%%%%%%%%%%%%%%%%%%%%%%%%%%%%%%%%%%%%%%%%%%%
%*************************************
\section{The Algorithm}\label{sec:the_algorithm}
The problem of self-consistency is solved, as anticipated in Section ~\ref{subsec:problems_nonlinear}, by alternating the solution of the TDSE ($VTE$) and the optimization of the potential ($Pot. Opt$).
The intrinsic nonlinear nature of the SP equation is reconciled with the requirements of a quantum circuit implementation imposing the correct normalization of the \textit{physical} wavefunction and potential, as given by  Eq.~\eqref{eq:physical_quantum_relation} and Eq.~\eqref{eq: potential structure}), respectively.
A scheme of this algorithm is reported in Alg.~\ref{alg: VTE scheme}, where $\{ \boldsymbol{\theta}_{t_i} \}$ and $\{ \boldsymbol{\phi}_{t_i} \}$  refer to the parameters' set at time $t_i;$ $i \in \{ 0, 1, ..., N_t-1  \}$. 
For conciseness, in Alg.~\ref{alg: VTE scheme} we use the following notation $\Psi_i \equiv \Psi(\boldsymbol{\theta}_{t_i}) $.
%%%%%%%%%%%%%%%%%%%%%%%%%%%%%%%%%%

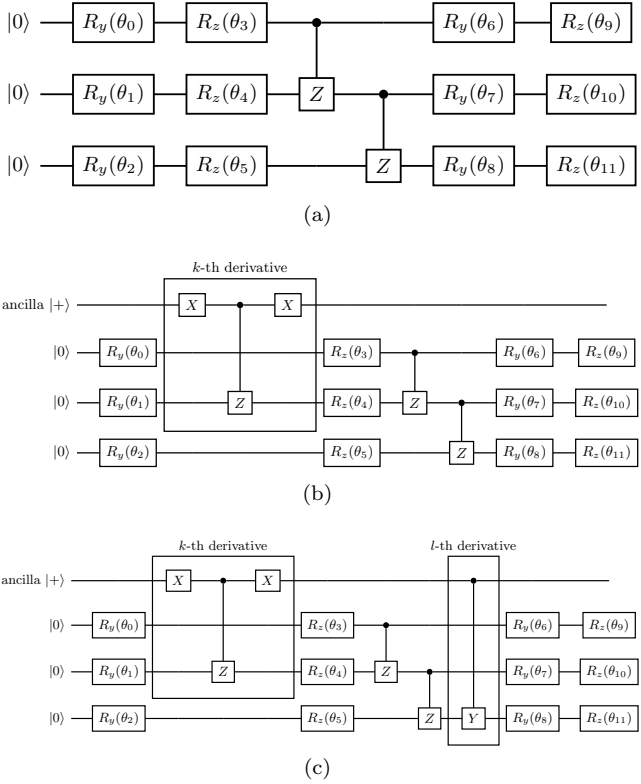
\begin{figure}
    \centering
    \subfloat[]{\resizebox{\columnwidth}{!}{
        \begin{quantikz}
            \lstick{$\ket{0}$} 
             &\gate{R_y(\theta_0)} &\gate{R_z(\theta_3)} &\ctrl{1} &\qw 
             &\gate{R_y(\theta_6)} &\gate{R_z(\theta_9)}
             \\
             \lstick{$\ket{0}$} 
             &\gate{R_y(\theta_1)} &\gate{R_z(\theta_4)} &\gate{Z} &\ctrl{1} 
             &\gate{R_y(\theta_7)} &\gate{R_z(\theta_{10})} 
             \\
             \lstick{$\ket{0}$} 
             &\gate{R_y(\theta_2)} &\gate{R_z(\theta_5)} &\qw &\gate{Z} 
             &\gate{R_y(\theta_8)} &\gate{R_z(\theta_{11})}
        \end{quantikz}
    }}%
    \\
    \subfloat[]{\resizebox{\columnwidth}{!}{
        \begin{quantikz}
                        \lstick{ancilla $\ket{+}$} &\qw
            &\gate{X} 
            \gategroup[wires=3,steps=3,style={inner sep=6pt}]{$k$-th derivative} 
            &\ctrl{2}   &\gate{X}
             &\qw&\qw&\qw 
             &\qw&\qw
             \\
             \lstick{$\ket{0}$} 
             &\gate{R_y(\theta_0)} &\qw &\qw &\qw &\gate{R_z(\theta_3)} &\ctrl{1} &\qw 
             &\gate{R_y(\theta_6)} &\gate{R_z(\theta_9)}
             \\
             \lstick{$\ket{0}$} 
             &\gate{R_y(\theta_1)} &\qw &\gate{Z} &\qw &\gate{R_z(\theta_4)} &\gate{Z} &\ctrl{1} 
             &\gate{R_y(\theta_7)} &\gate{R_z(\theta_{10})} 
             \\
             \lstick{$\ket{0}$} 
             &\gate{R_y(\theta_2)} &\qw &\qw &\qw &\gate{R_z(\theta_5)} &\qw &\gate{Z} 
             &\gate{R_y(\theta_8)} &\gate{R_z(\theta_{11})}
        \end{quantikz}
    }}%
    \\
    \subfloat[]{\resizebox{\columnwidth}{!}{
        \begin{quantikz}
            \lstick{ancilla $\ket{+}$} &\qw
            &\gate{X} 
            \gategroup[wires=3,steps=3,style={inner sep=6pt}]{$k$-th derivative} 
            &\ctrl{2}   &\gate{X}
             &\qw&\qw&\qw 
             &\ctrl{3}
             \gategroup[wires=4,steps=1,style={inner sep=6pt}]{$l$-th derivative} 
             &\qw&\qw
             \\
             \lstick{$\ket{0}$} 
             &\gate{R_y(\theta_0)} &\qw &\qw &\qw &\gate{R_z(\theta_3)} &\ctrl{1} &\qw &\qw
             &\gate{R_y(\theta_6)} &\gate{R_z(\theta_9)}
             \\
             \lstick{$\ket{0}$} 
             &\gate{R_y(\theta_1)} &\qw &\gate{Z} &\qw &\gate{R_z(\theta_4)} &\gate{Z} &\ctrl{1} &\qw 
             &\gate{R_y(\theta_7)} &\gate{R_z(\theta_{10})} 
             \\
             \lstick{$\ket{0}$} 
             &\gate{R_y(\theta_2)} &\qw &\qw &\qw &\gate{R_z(\theta_5)} &\qw &\gate{Z} &\gate{Y}
             &\gate{R_y(\theta_8)} &\gate{R_z(\theta_{11})}
        \end{quantikz}
    }}%
\caption{ 
    Quantum circuits used to prepare: $(a)$ the trial wavefunction; $(b)$ the unitary matrix $F_k$ that generates states like the one in Eq~\eqref{eq: specific  structure of states for mixed}; $(c)$ the unitary matrix $F_{k, l}$ that generates states like the one in Eq.~\eqref{eq: specific  structure of double mixed}. 
}\label{fig: derivatives circuits}
\end{figure}

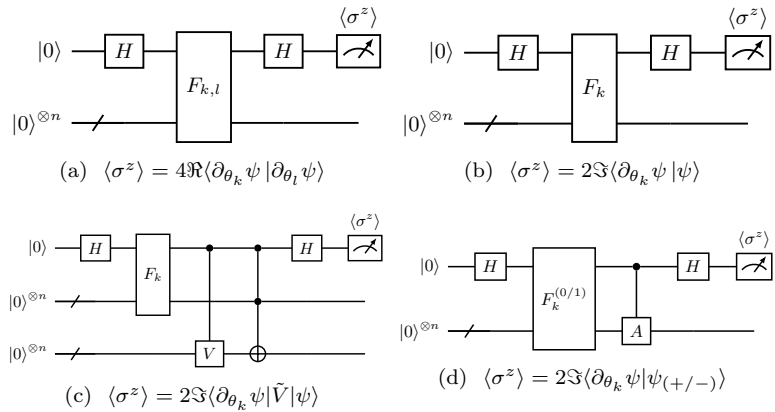
\begin{figure*}[t]
    \centering
    %\hspace{3cm}
    \subfloat[\label{fig: imag superpos}
        $ \langle \sigma^z \rangle  
        =
        4\Re
        \langle \partial_{\theta_k} \psi \ket{\partial_{\theta_l}\psi}$
    %]{\resizebox{0.48\columnwidth}{!}{
    ]{\resizebox{0.6\columnwidth}{!}{
        \begin{quantikz}
             \lstick{$\ket{0}$} &\gate{H} &\gate[wires = 2]{F_{k,l}} &\gate{H} &\meter{$\langle \sigma^z \rangle$} 
             \\
             \lstick{$\ket{0}^{\otimes n}$} &\qw\qwbundle{} & &\qw&\qw 
        \end{quantikz}
    }}%
    %\hfill
    \subfloat[\label{fig: real superpos}
    $ \langle \sigma^z \rangle  
        = 2\Im
        \langle \partial_{\theta_k} \psi \ket{\psi}$
    %]{\resizebox{0.48\columnwidth}{!}{
    ]{\resizebox{0.6\columnwidth}{!}{
        \begin{quantikz}
             \lstick{$\ket{0}$} &\gate{H} &\gate[wires = 2]{F_{k}} &\gate{H} &\meter{$\langle \sigma^z \rangle$} 
             \\
             \lstick{$\ket{0}^{\otimes n}$} &\qw\qwbundle{} & &\qw &\qw 
        \end{quantikz} 
    }}%
    \\
    \subfloat[\label{fig: new mixed potential}
    $\langle \sigma^z \rangle = 
    2 \Im \langle \partial_{\theta_k} \psi| \Tilde{V} | \psi \rangle$ 
    %]{\resizebox{0.48\columnwidth}{!}{
    ]{\resizebox{0.6\columnwidth}{!}{
        \begin{quantikz}
                 \lstick{$\ket{0}$} &\gate{H} &\gate[wires=2]{F_k} &\ctrl{2} &\ctrl{1} &\gate{H} &\meter{$\langle \sigma^z \rangle$} 
                 \\
                 \lstick{$\ket{0}^{\otimes n}$} &\qw\qwbundle{} & &\qw &\ctrl{1} &\qw &\qw  \\
                 \lstick{$\ket{0}^{\otimes n}$} &\qw\qwbundle{} &\qw &\gate{V} &\targ{} &\qw &\qw
        \end{quantikz}  
    }}
    %\hfill
    \subfloat[\label{fig: new mixed kinetic}
    $\langle \sigma^z \rangle = 
    2 \Im \langle \partial_{\theta_k} \psi| \psi_{(+/-)} \rangle$ 
    %]{\resizebox{0.48\columnwidth}{!}{
    ]{\resizebox{0.6\columnwidth}{!}{
        \begin{quantikz}
                 \lstick{$\ket{0}$} &\gate{H} &\gate[wires = 2]{F^{(0/1)}_{k}} &\ctrl{1} &\gate{H} &\meter{$\langle\sigma^z\rangle$} 
                 \\
                 \lstick{$\ket{0}^{\otimes n}$} &\qw\qwbundle{} & &\gate{A}  &\qw &\qw
            \end{quantikz}
    }}
\caption{ 
    Quantum circuits for the evaluation of the VTE matrix elements in Eq.~\eqref{eq: McLach M}, \eqref{eq: McLach V} through the measurement of the ancilla qubits $\langle \sigma_z \rangle$. 
    The correspondence between expectation value and measurement is reported under the respective scheme.
    The unitaries $F_{k,l}$ and $F_k$ are both reported in Fig.~\ref{fig: derivatives circuits}. They are used to produce the wavefunction and its derivatives.
    The Toffoli gate in panel (c) represents a Toffoli ladder: $n$ Toffoli gates linking the wavefunction and the potential register qubit per qubit.
    (d)  $F_k^{(1/0)}$ denotes $F_k$ with different control states ($\ket{0}$ or $\ket{1})$ and $A$ is the adder circuit~\cite{lubasch_variational_2020} (See Appendix~\ref{sec: supp: adder} 
    %Supplemental Material at [URL will be inserted by publisher] 
    for more details on the adder circuit).
}\label{fig: new circuit hamiltonian}
\end{figure*}
%

%-----------------------------------------------------------
%//////////////////////////////////////////////////////////
%
\subsection{Circuit implementation}\label{subsec: circuit iplementation}

The trial quantum states for both the wavefunction and the potential are implemented using a heuristic local ansatz~\cite{ollitrault_non-adiabatic_2020,ollitrault_quantum_2022, lubasch_variational_2020, mocz_towards_2021, cerezo2021variational, chakrabarti_threshold_2021}
that alternates single qubit rotation layers $U^{rot}(\boldsymbol{\theta})$ and entangling layers $U^{ent}$ (see example in Fig. \ref{fig:ry_cnot})
\begin{equation} 
    \label{eq: ansatz_structure}
    U(\boldsymbol{\theta})
    =
    U_0^{rot}(\boldsymbol{\theta}_0)
    \cdot
    \prod_{\xi = 1}^{D}  
    U_\xi^{ent} \cdot
    U_\xi^{rot} (\boldsymbol{\theta}_\xi)
\end{equation}
where $D$ is the number of entangling layers and $\boldsymbol{\theta}_\xi$ a subgroup of parameters.
In Fig~\ref{fig: derivatives circuits}, we show the typical circuits used to encode the wavefunction $\ket{\psi(\boldsymbol{\theta})}$,
while Fig.~\ref{fig:ry_cnot} reports the one used for the potential $\ket{\Phi_{\Tilde{V}}(\boldsymbol{\phi})}$.
The latter consists of just $\mathrm{R}_Y(\theta)$ rotations and $\mathrm{CX}$ gates, since the target potential function is real-valued.

%%%%%%%%%%%%%%%% Parlo velocemente di come calcolare le derivate: schema e matrice W unitaria.
The quantum part of the evolution algorithm resides in the measurement of the expectation values in Eqs.~\eqref{eq: McLach M} and \eqref{eq: McLach V}. 
In the following, we propose an efficient implementation of the circuits for the evaluation of the terms with derivatives in Eqs.~\eqref{eq: McLach M} and \eqref{eq: McLach V}. 
In particular, we provide a detailed procedure for the calculation of those matrix elements that have a functional dependence on the non-linear potential, such as 
$\bra{\partial_{\theta_k}\psi} \mathcal{H}(V(\boldsymbol{\phi})) \ket{\psi}$. 

%%%%%%%%%%%%%%%% Schema e spiegazione dell'improvement metodo F..  
Given the structure of the ansatz in Eq.~\eqref{eq: ansatz_structure} 
and $\theta_k$ in the subset $\boldsymbol{\theta}_{\tilde{\xi}}$
, the derivatives $\partial_{\theta_k}$ 
leaves the unitary unchanged, with the exception of the target rotational layer:
\begin{equation} \label{eq: rotational_layer}    
    U_{\Tilde{\xi}}^{rot}(\boldsymbol{\theta}_{\Tilde{\xi}})
    = \bigotimes_{j=0}^{n-1} \exp \{-\frac{i}{2} \alpha_j \theta_{\tilde{\xi},j} \}
\end{equation}
where $\theta_{\tilde{\xi}, j} \in \boldsymbol{\theta}_{\tilde{\xi}}$ and $\alpha_j \in \{X, Y, Z \}$ is a Pauli matrix, generator of single qubit rotations.
Combining Eqs.~\eqref{eq: ansatz_structure}, \eqref{eq: rotational_layer}
and 
${\ket{\psi(\boldsymbol{\theta})} = U(\boldsymbol{\theta})\ket{\boldsymbol{\Xi}}}$,
one gets for the partial derivative $\partial_k$ 
\begin{equation}\label{eq:derivative}
    \partial_{\theta_k} U(\boldsymbol{\theta})\ket{\Xi}
    = \ket{\partial_{\theta_k} \psi(\boldsymbol{\theta})} = -\frac{i}{2} W_k(\boldsymbol{\theta}) \ket{\Xi} \, ,
\end{equation}
for a generic quantum state $\ket{\mathbf{\Xi}}$. 
Here 
$W_k(\boldsymbol{\theta})$ 
is a modified version of $U(\boldsymbol{\theta})$ where the single qubit rotation $R_\alpha(\theta_k)$ is preceded by its own generator~\cite{ollitrault_quantum_2022, schuld_evaluating_2019}.
\\
In the search for an efficient quantum circuit able to reproduce the matrix and vector elements of the McLachlan equation of motion of Eq. \eqref{eq: McLach eq. of motion}, the main obstacle is to produce a quantum state with the following structure:
\begin{equation}\label{eq: structure of states for mixed}
    \ket{\psi} = \frac{1}{\sqrt{2}}\left( U_1(\boldsymbol{\theta})\ket{\mathbf{\Xi}}\ket{0} + U_2(\boldsymbol{\theta})\ket{\mathbf{\Xi}}\ket{1} \right)\,,
\end{equation}
where $U_1, U_2$ are generic unitaries and the second quantum register (single qubit) is used to evaluate the value of the matrix element. 
In the specific case at study, these unitaries should be expressive enough to enable a suitable parameterization of the wavefunction and its derivatives (Eq.~\eqref{eq:derivative}).
Given the structure of the circuit $W_k$, 
by controlling only the Pauli matrix that implements the derivative, it is possible to prepare the quantum states 
\begin{align}
    \small
    F_k (\boldsymbol{\theta})
    &
    \ket{\mathbf\Xi}\ket{+} 
    \nonumber
    \\
    \label{eq: specific  structure of states for mixed}
    & =
    \nonumber
    \frac{1}{\sqrt{2}}\left( W_k(\boldsymbol{\theta})\ket{\mathbf{\Xi}}\ket{0} + U(\boldsymbol{\theta})\ket{\mathbf{\Xi}}\ket{1} \right)
    \\
    & =
    \frac{1}{\sqrt{2}}
    \left( 2i
    \ket{\partial_{\theta_k}\psi(\boldsymbol{\theta})}\ket{0} + \ket{\psi(\boldsymbol{\theta})}\ket{1}
    \right),
    \\
    \nonumber
    F_{k, l} (\boldsymbol{\theta})
    &
    \ket{\mathbf{\Xi}}\ket{+} 
    \\
    \label{eq: specific  structure of double mixed}
    & =
    \nonumber
    \frac{1}{\sqrt{2}}\left( W_k(\boldsymbol{\theta})\ket{\mathbf{\Xi}}\ket{0} + W_l(\boldsymbol{\theta})\ket{\mathbf{\Xi}}\ket{1} \right)
    \\
    & =
     i\sqrt{2}
    \Big( 
    \ket{\partial_{\theta_k}\psi(\boldsymbol{\theta})}\ket{0} + 
    \ket{\partial_{\theta_l}\psi(\boldsymbol{\theta})} \ket{1}
    \Big) \, ,
\end{align}
for a given reference state $\ket{\mathbf\Xi}$ where $F_{k, l}(\boldsymbol{\theta})$ and $F_k(\boldsymbol{\theta})$ refer to unitaries for the different derivatives (see Fig.~\ref{fig: derivatives circuits}).

Fig. \ref{fig: new circuit hamiltonian} summarizes 
all quantum circuits relevant for the evaluation of the terms 
in Eq. \eqref{eq: McLach M}, \eqref{eq: McLach V}. 
A brief discussion on how to evaluate them on a QC will follow, starting with the overlaps
$\Im\bra{\partial_{\theta_k}\psi} \psi\rangle$ and 
$\Re\bra{\partial_{\theta_k}\psi} \partial_{\theta_j}\psi\rangle$.
One can notice from Eqs.\eqref{eq: specific  structure of states for mixed},~\eqref{eq: specific  structure of double mixed} that upon applying a $H$ gate, measuring $\langle\langle\sigma_z\rangle\rangle$ on the ancillary qubit returns the desired quantities.
Furthermore, there is no need to evaluate the real part to compute the product of the overlaps in Eq.~\eqref{eq: McLach M} since the term $\bra{\partial_{\theta_k}\psi} \psi\rangle$ is purely imaginary.
The circuits used to do so are shown in Figs.~\ref{fig: imag superpos},~\ref{fig: real superpos}. 

The potential part  
$\Im \braket{\partial_{\theta_{k}} \psi | \Tilde{V}(\tilde{\boldsymbol{\phi}}) | \psi}$, 
is what actually connects the solution of the $TDSE$ and the Poisson equation.  
$\Tilde{V}(\tilde{\boldsymbol{\phi}})$ is given  in Eq. (\ref{eq: potential structure}) and is prepared using the parameters $\tilde{\boldsymbol{\phi}}$ resulting from the minimization of Eq.~(\ref{eq:analytic_cost_potential}). 
The circuit in Fig.~\ref{fig: new mixed potential}
is the one used for the evaluation of this linking term, where the series of $n$ Toffoli gates provides a pointwise multiplication between the wavefunction and the potential registers (i.e, $\sum_k \tilde{V}_k \psi_k)$. 
% ----------------------------

Concerning the term 
$\Im 
\left\{
\bra{\partial_{\theta_k} \psi} \nabla^2 \ket{\psi}
\right\}$, a few considerations are needed. For systems of cosmological relevance, we expect accurate simulations to require a fine enough spatial resolution to resolve all spatial features. 
Therefore, using a finite differences approach, as also proposed in Ref.~\cite{lubasch_variational_2020}, can be justified as the discretization error should be irrelevant at higher resolutions. 
In this framework, an approximation of the Laplace operator is given by 
\begin{equation}\label{eq: mixed finite diff kinetic}
    \begin{aligned}
    \Im 
    &
    \left\{
    \bra{\partial_{\theta_k} \psi} \nabla^2 \ket{\psi}
    \right\}
    = \\
     &
    \frac{1}{\Delta x^2}\Im \left\{
    \bra{\partial_{\theta_k} \psi} \psi_+ \rangle
    - 2 \bra{\partial_{\theta_k} \psi} \psi \rangle
    + \bra{\partial_{\theta_k} \psi} \psi_{-} \rangle
    \right\}\,,
    \end{aligned}
\end{equation}
with the positive (and negative) shifted wavefunctions
${\ket{\psi_\pm} = \sum_{j=0}^{N-1} \psi_{j \pm 1} \ket{\mathrm{bin}(j)}}$, 
obtained using the adder circuit $A$~\cite{lubasch_variational_2020}, whose action on the $j$-th base is
${\ket{\mathrm{bin}(j)} \mapsto \ket{\mathrm{bin}(j-1)}}$, in combination with the unitary $F_k(\boldsymbol{\theta})$ of Eq.~\eqref{eq: specific  structure of states for mixed}
with different control state allows to evaluate the shifted overlaps in Eq.~\eqref{eq: mixed finite diff kinetic}. A scheme of the circuits needed to perform these operations is presented in Fig.~\ref{fig: new mixed kinetic}.

See Appendix~\ref{sec: supp: circuit proofs}
%Supplemental Material 
for more details about the functioning the circuits in Figs.~\ref{fig: new mixed kinetic},~\ref{fig: new mixed potential}.
%
% --------------------------------------------------
%
%%%%%%%%%%%%%%%%%%%%%%%%%%%%%%%%%%%
%

%------------------------------------

\begin{table*}[]
\caption{ 
    State fidelity $\mathcal{F}$ (between the classical reference and the evolved state at $t=3$) for different VTE simulations. Hyperparameters: 
    $D_{\psi}$ and $D_{V}$, number of rotation layers in the wavefunction and potential ansatz respectively; $M_p$ total number of parameters in the wavefunction ansatz;
    $N_t$ number of timestep used in the simulation; $r_c$ cutoff for singular values, used determine the effective rank of the matrix $\mathrm{M}$ in Eq.~\eqref{eq: McLach eq. of motion} (more information available in \texttt{Scipy} documentation~\cite{scipy_lstsq}); $\epsilon$ regularization factor added to the diagonal of the matrix $\mathrm{M}$ in Eq.~\eqref{eq: McLach M}.
    }
    \label{tab: hyperparameters}
    \centering
    \begin{tabular}{p{2.5cm} p{1.5cm} p{1.5cm}  p{1.5cm}  p{1.5cm}  p{1.5cm}  p{1.5cm}  p{1cm} }
         \hline \hline
         &$D_{\psi}$ &$D_{V}$ &$M_p$ &$N_t$ &$r_c$ &$\epsilon$
         &$\mathcal{F}$
         \\
         \hline
         $4$-qubits &$4$ &$4$ &$32$ 
         &$6\cdot 10^2$ &$10^{-7}$ &$10^{-3}$
         &$0.976$
         \\
         $5$-qubits &$5$ &$6$ &$50$
         &$9\cdot10^3$ &$10^{-8}$ &$10^{-4}$
         &$0.944$
          \\
          $5$-qubits &$5$ &$6$ &$50$
         &$2\cdot10^4$ &$10^{-8}$ &$10^{-4}$
         &$0.960$
         \\
          $5$-qubits &$6$ &$6$ &$60$
         &$6\cdot10^3$ &$10^{-8}$ &$10^{-4}$
         &$0.956$
          \\
          \hline \hline
    \end{tabular}
\end{table*}
%

%---------------
\begin{figure*}
    \begin{tikzpicture}
        \node at (0,0) {\includegraphics[width = 2\columnwidth]{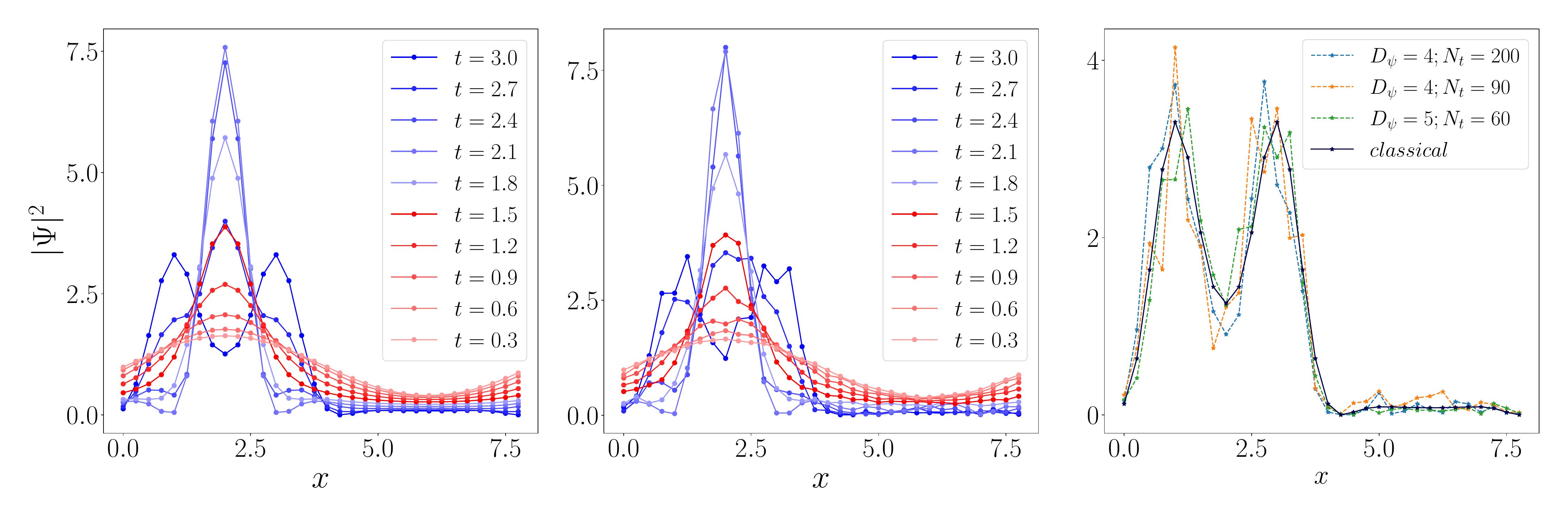}};
        \node at (0.4, -2.9) {(b)};
        \node at (-5.1, -2.9) {(a)};
        \node at (5.95, -2.9) {(c)};
    \end{tikzpicture}
\caption{ 
Comparison between probability distributions at different times for a 5 qubits system and $\lambda=1$. The left panel (a) is the classical reference, obtained with a spectral method~\cite{mocz_towards_2021}. In the middle one (b) are presented the results obtained through a VTE simulation (using the algorithm in Fig.~\ref{alg: VTE scheme}). In the left (c) we compare the classical probability distribution at $t=3$ with the results obtained from the VTE simulations with different hyperparameters (more details in Tab.~\ref{tab: hyperparameters}) . The one chosen  for the simulation in the middle panel (b) are $N_t = 2 \cdot 10^4$, $\epsilon = 10^{-4}$, $r_c = 10^{-8}, D_\psi = 6, D_V = 6$. }.
\label{fig: 5qb_VTE}
\end{figure*}
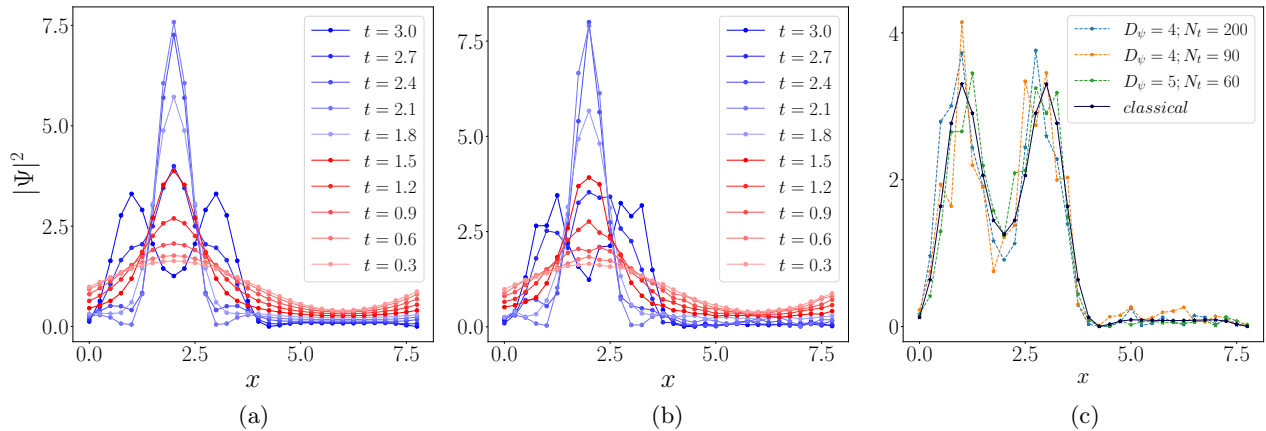

%---------------

\section{Results and Discussion}\label{sec:results}

Before addressing the setup used in our simulation, some consideration about the characteristic scales appearing in the SP equation and the corresponding units are needed.

Given the invariance of the SP Eqs~\eqref{eq:Sch_equation_our},~\eqref{eq:poisson_equation} under the following scaling transformation:
\begin{equation}
    \label{eq: scaling tranformation}
    \{ x, t, \psi, \lambda\} \mapsto 
    \{ \alpha x, \beta t, \beta^{-1} \psi, \alpha^{-2} \beta\lambda\} \, 
\end{equation}
$\lambda$ emerges as an intrinsic scale of the problem~\cite{mocz_schrodinger-poissonvlasov-poisson_2018} as its scaling law combines changes in both the spatial and time domain (i,e system with different box dimension or evolution time will display different dynamics). 

Concerning the dimension of the physical quantities appearing in the problem, we used arbitrary unit. 
See Appendix~\ref{supp: SP rescaling} 
%Supplemental Material at [URL will be inserted by publisher] 
for details on the arbitrary values chosen for the density normalization $\rho^*$ and the constant $G$ in the transition from Eq.~\eqref{eq:fuzzy_SP} to Eqs.\eqref{eq:Sch_equation_our},~\eqref{eq:poisson_equation}.
%%%%%%%%%%%%%%%%%%

As a final remark, we would like to emphasize that in this preliminary study all simulations were performed in an idealized setting, without considering gate errors and sampling shot noise.
%%%%%%%%%%%%%%%%%%
%------------------------------

\subsection{Numerical simulations}\label{subsec:Numerical simulations}
We consider a one dimensional system of length $L=8$ with  periodic boundary conditions. 
As anticipated above, we use arbitrary units for both spatial coordinates and time variable.
The choice of $L$ and the total time of the simulation is done in such a way that,
once we fix $\lambda = 1$, the self interacting potential Eq.~\eqref{eq:poisson_equation} exactly balances the diffusion associated to the Schr\"odinger time-evolution.
In order to compare our results with those from Ref.~\cite{mocz_towards_2021}, we used as initial condition a sinusoidal
distribution of the form
\begin{equation}\label{eq: initial condition}
    \Psi(x, 0) = \sqrt{ 1 + 0.6 \sin{\left(\frac{\pi}{4} x \right)}} \, ,
\end{equation}
evolved according to Eqs.~\eqref{eq:Sch_equation_our} and~\eqref{eq:poisson_equation}.
This specific initial condition is a well-known standard test case. It corresponds to one of the different Fourier components typically found in initial distributions for the VP equations, like Gaussian random fields. It is widely used as it makes it easy to observe the effects of shell-crossing. We will delve deeper into the concept of shell crossing in Sect.~\ref{subsec:Interpretation of sp res} for further clarification.

For this proof-of-principle numerical implementation, the parameters $\boldsymbol{\theta_0}$ reproducing the initial quantum state are obtained by optimizing the state fidelity $\mathcal{F}(\psi(\boldsymbol{\theta}), \tilde{\psi})$ between the variational trial state $\ket{\psi(\boldsymbol{\theta})}$ and a target state $\ket{\tilde{\psi}}$.
In this work we refer to $\mathcal{F}$ as the state fidelity between two \textit{quantum} states~\cite{state_fidelity} (i.e, state normalization is $1$). 
In the situation where $\ket{\psi_1}$ and $\ket{\psi_2}$ are pure states, we have
${
    \mathcal{F}(\psi_1, \psi_2) = |\langle \psi_1 | \psi_2\rangle|^2
}$.
This value will be also used to measure of the convergence of the states obtained with the variational method to the ones obtained classically. We point out that this has noting to do with the convergence to the actual solution of the physical problem (i.e, does not take into consideration the grid discretization error). 
The classical optimization of the potential (
\textit{Pot. Opt.} in Algorithm~\ref{alg: VTE scheme}
) is performed using a combination of \texttt{COBYLA} (to start the optimization) and \texttt{BFGS} (to find the best solution) algorithms as implemented in \texttt{SCIPY} v1.9.0.
All simulations were performed in Qiskit~\cite{qiskit} within the \texttt{statevector} framework, i.e., using a matrix representation of the quantum circuit and a vector representation of the quantum state.

The equations of motion in Eq.~\eqref{eq: McLach eq. of motion} are integrated using an explicit Euler method with fixed timestep for a total of $N_t$ steps. Here, it is important to mention that, in general, the inversion of the matrix $\mathrm{M}$ in Eq.~\eqref{eq: McLach M} may become ill-defined. To reduce the resulting instabilities of the dynamics, we used the \texttt{SCIPY} least squares solver \cite{scipy_lstsq} with a suitable choice of the corresponding hyperparameters:
the cutoff $r_c$, 
used to determine the effective rank of the matrix in Eq.~\eqref{eq: McLach eq. of motion} such that the singular values smaller than $r_c \cdot \Lambda_{max}$ are set to zero (here $\Lambda_{max}$ is the singular value of largest magnitude), and the regularization factor $\epsilon$, applied to the diagonal of the matrix $\mathrm{M}$ in Eq.~\eqref{eq: McLach M}.

In order to determine the quality of the results, we should also consider the level of expressivity of the variational ansatz, which is used to encode the system wavefunction and the potential. In order to achieve accurate results, one would need – in principle -  a number of circuit parameters $\boldsymbol\theta(t)$ for the wavefunction that approaches the size of the Hilbert space. On the other hand, the number of terms in the matrices and vectors used in the equations of motion, Eqs.~\eqref{eq: McLach M} and ~\eqref{eq: McLach V}, scale as $M_p^2$ and $M_p$, respectively, as shown in Tab.~\ref{tab: different_terms}, where $M_p$ is the number of parameters. Reducing the number of parameters significantly reduces the total number of circuit evaluations. This however translates in a lower accuracy of the dynamics, as the ansatz may not enable a thoroughly description of the sector of interest of the full Hilbert space. 
Similarly, a large number of parameters will enable a more accurate description of the self-consistent potential, at the price of a more cumbersome (classical) optimization process and an increased circuit depth. 
%--------------------------

%-------------------
To assess the quality of our implementation (including the adjustment of the hyperparameters), we performed two series of simulations.
The first one is a \textit{classical spectral method} based on the $FFT$ as in~\cite{mocz_towards_2021}.
Results obtained from this approach will be used as a reference.
The actual implementation of our proposed quantum algorithm %(as one would execute on quantum hardware) 
consists, instead, of repeated cycles of circuit optimization and VTE steps (Algorithm~\ref{alg: VTE scheme}). When comparing its outcomes with the exact ones (Fig.~\ref{fig: 5qb_VTE} and Tab.~\ref{tab: hyperparameters}) we observe that the quantum approach rightfully captures the qualitative behaviour of the wavefunction, although the probability distribution obtained from the VTE is not as smooth as the exact one.

%%%%%%%%%%%%%%%%%%%%%%%%%%%%%%%%%%%
\begin{table}[t]
    \centering
    \caption{Number of different circuits used to evaluate the terms in Eq.~\eqref{eq: McLach eq. of motion} with the respective number of qubits needed for the implementation. Here $M_p$ is the number of variational parameter in the wavefunction ansatz and $n=log_2N$ the number of qubits used for the discretization.}
    \label{tab: different_terms}
    \begin{tabular}{p{3cm} p{3cm} p{2cm}}
        \hline \hline
         Term &No. circuits &No. qubits 
         \\ \hline
        $\Re \braket{\partial_{\theta_k}\psi | \partial_{\theta_l}\psi}$  &$M_p(M_p-1)/2$  &$n+1$
        \\
        $\Im \braket{\partial_{\theta_k}\psi | \psi}$  &$M_p$  &$n+1$
        \\
        $\Im \braket{\partial_{\theta_k}\psi | \tilde{V} | \psi}$  &$M_p$  &$2n+1$
        \\
        $\Im \braket{\partial_{\theta_k}\psi | \psi_\pm}$  &$2M_p$  &$2n-1$
        \\ \hline \hline
    \end{tabular}
    \end{table}

%-------------------------

%%%%%%%%%%%%%%%%
%%%%%%%%%%%%%%%%%%%%%%%
%%%%%%%%%%%%%%%%%%%%%%%%%%%%%%%%%%%%%%
%%%%%%%%%%%%%%%%%%%%%%%
%%%%%%%%%%%%%%%%
\subsection{Interpretation of the SP results}
\label{subsec:Interpretation of sp res}
\begin{figure}
    \begin{tikzpicture}
        \node at (0,0) {\includegraphics[width = \columnwidth]{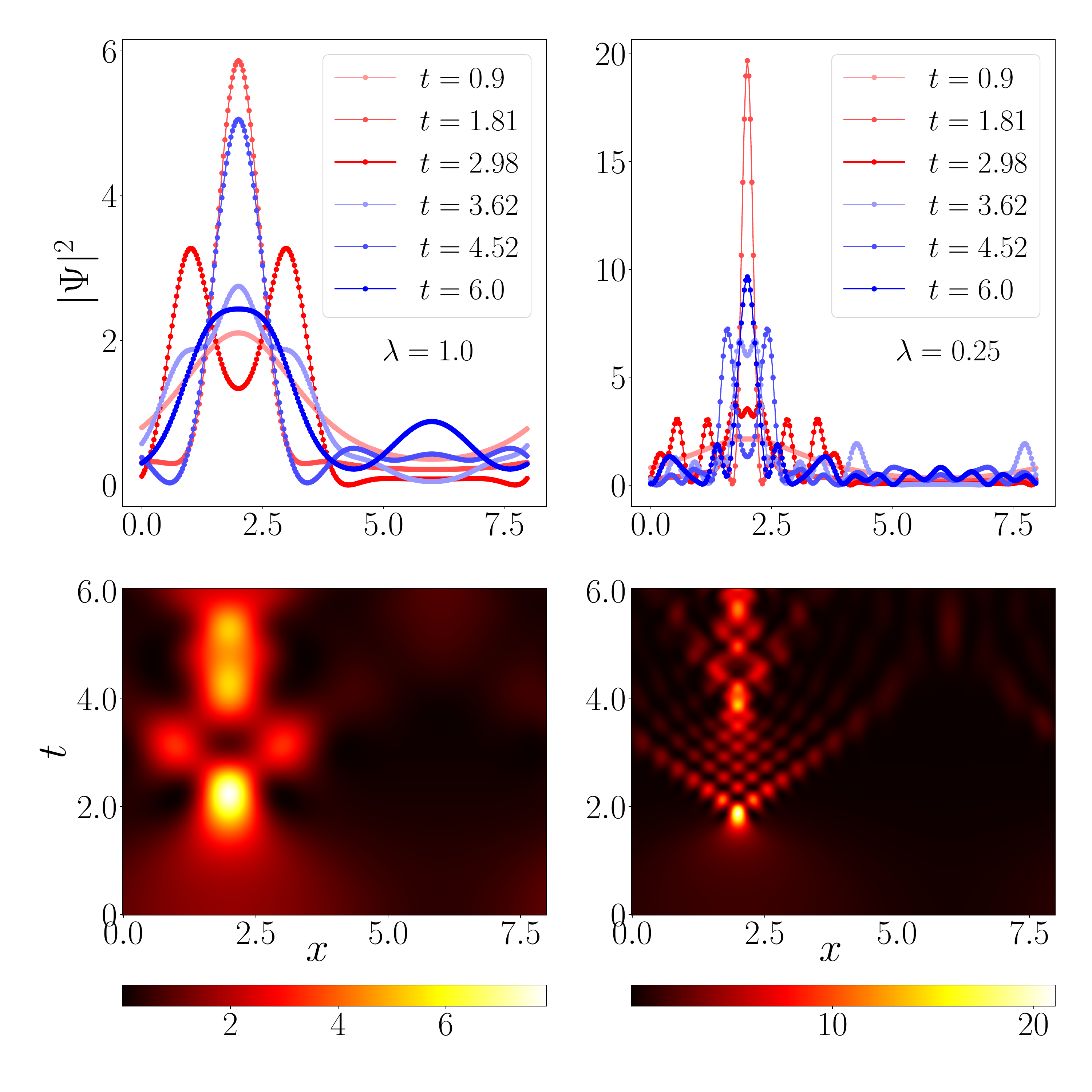}};
        \node at (-1.65, -4.4) {(a)};
        \node at (2.3, -4.4) {(b)};        
    \end{tikzpicture}
\caption{ Classical evolution of the $1$-D probability distribution under the effect of the gravitational potential, for different values of $\lambda$. Both simulations have been carried out with a spectral method \cite{mocz_towards_2021}. In the top row, probability distributions are shown at fixed time frames. In the bottom row the same evolution is shown in $2$-D perspective by a heatplot: the $x$ axis represents the spatial coordinate, while the $y$ axis is used for the time; the probability distribution magnitude is represented through a color gradient. The difference between these two simulations is given by the intensity of the quantum pressure term. In the left column (a) $\lambda=1$ and the quantum effect balances the diffusion; In the right column (b), with $\lambda = 1/4$ the dynamics is similar to the classical  one (VP).}
\label{fig: different lambda plots}
\end{figure}

%---- DISCUSSIONE DELLE SIMULAZIONI CON LAMBDA = 1 QUANUTM!
Fig.~\ref{fig: different lambda plots} shows the time evolution of the initial sinusoidal distribution, as given in  Eq.~\eqref{eq: initial condition}, over a time span of approximately $6$ time units for two different choices of the parameter $\lambda$ 
(left: $\lambda=1$, right: $\lambda=0.25$). The lower panels depict the same dynamics as a two-dimensional surface plot of the time dependent wavefunction. The larger the value of $\lambda$, the larger the quantum nature of the dynamics; in fact, in the limit of $\lambda\rightarrow 0$, the SP dynamics converges towards the classical VP dynamics~\cite{mocz_schrodinger-poissonvlasov-poisson_2018}. 
Physically, the collapse and splitting of the probability distribution (left panels in Fig.~\ref{fig: different lambda plots}) is an effect of the self-interacting potential. This is regulated by the scale of the problem $\lambda$. 
However, as stated in the preamble of Sect.~\ref{sec:results}, what really matters is not the absolute value of $\lambda$, but its value relative to the box size and  time (e.g, if instead of $L=8$ we had $L=1$, we would need to change $\lambda$ to $\lambda/64$, accordingly). 
In the classical limit $\hbar/m \rightarrow 0$, the quantum effects are 
suppressed, the potential cannot counter anymore the diffusion and secondary peaks arise, as in the classical VP solution. 
In this scenario, the effects of shell-crossing are more pronounced. 
%

%\textcolor{blue}{
The term \textit{shell-crossing} can be better understood in the context of the study of the collapse of a spherical density perturbation in a self-gravitating collisionless fluid~\cite{shell-crossing-1}. Following an accretion due to the expansion of the Universe, spherical shells of matter collapse under the influence of gravity, until they intersect and a singularity arises. Subsequently, the term has been repurposed in the context of dark matter~\cite{shell-crossing2, shell-crossing3} due to its non-collisional nature.
More in general the shell-crossing happens whenever whenever the orbits of two, or more, fluid elements intersect. 
%}

%
An example is shown in the right column of Fig.~\ref{fig: different lambda plots}. Starting from the initial sinusoidal distribution, the gravitational attraction induces the concentration of the matter density in a first peak (around time $t=3$), which then collapses by effect of gravity damping. This process repeats few more times, giving rise to a multitude of sub-peaks as a result of repeated episodes of shell-crossing.

%----------------------------------
%%%%%%%%%%%%%%%%%%%%%%%%%%%%%%%%%%%
%%%%%%%%%%%%%%%%%%%%%%%%%%%%%%%%%%%%%%%  RESOURCE SCALING

\begin{figure*}
\centering
   \begin{tikzpicture}
        \node at (0,0) {\includegraphics[width = 1.7\columnwidth]{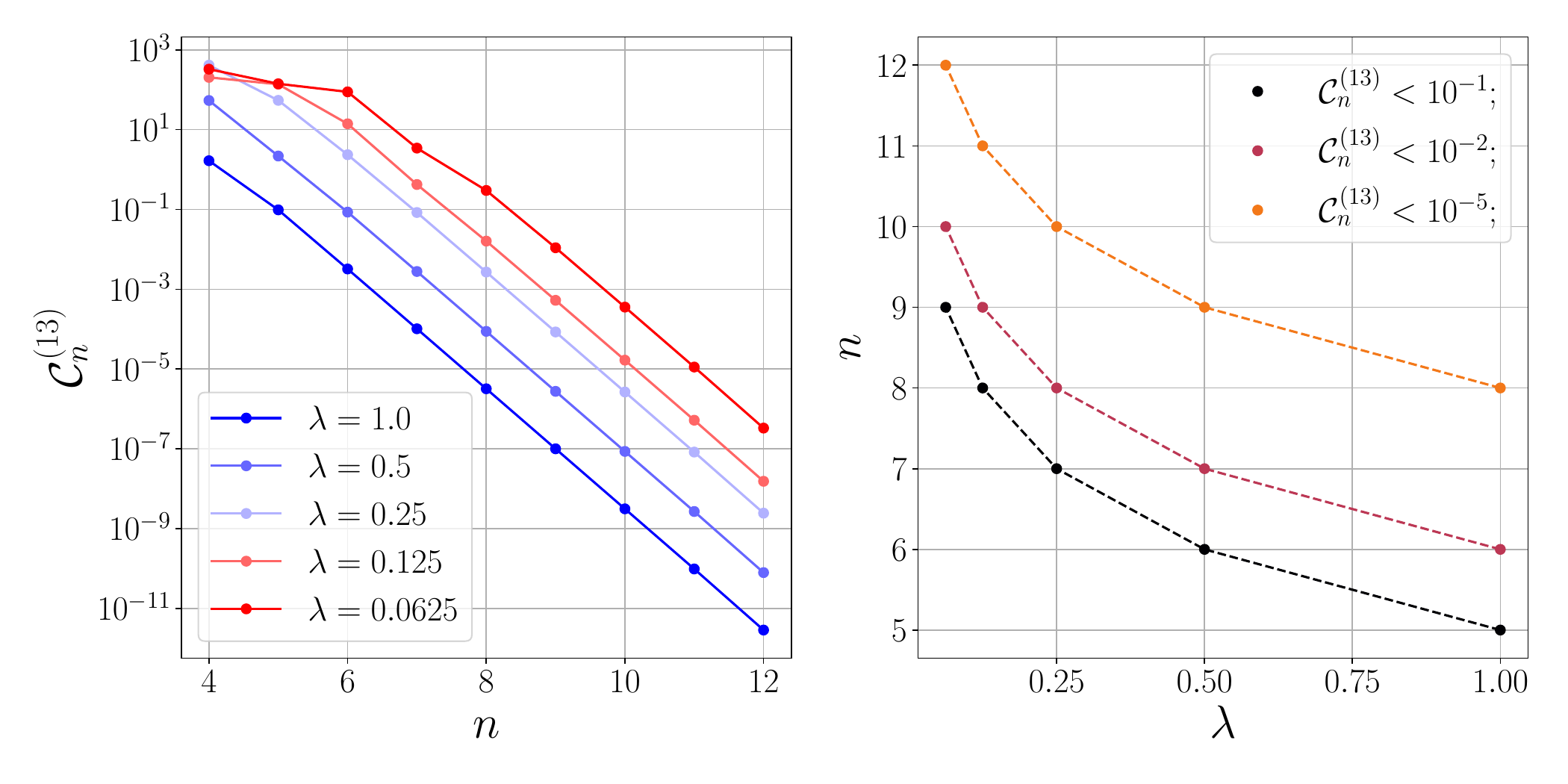}};
        \node at (-2.9, -3.7) {(a)};
        \node at (4.1, -3.7) {(b)};
    \end{tikzpicture}
\caption{ 
(a) Convergence $\mathcal{C}^{(13)}_n$ (see Eq.~\eqref{def_mathcalC}) as a function of the resolution, given by the number of qubits $n$, for different values of $\lambda$ (scale for quantum effects).
To match the number of points between the two systems, extra points are taken onto the connecting line between two adjacent points in the $n$ qubits discretization.
(b) Minimum number of qubits required to obtain a fixed arbitrary fidelity $\tilde{\mathcal{C}}^{(13)}$ as a function of $\lambda$.
}
\label{fig: scaling_lambda}
\end{figure*}

%---------------

%\subsection{Resources scaling}
\subsection{Scaling of required resources}
\label{subsec:resource scaling}
The largest cosmological simulations describe nowadays the evolution of boxes having a size of several Gigaparsecs, and using of the order of a trillion resolution elements (particles) \cite{Ishiyama.etal.2021}. While simulations of this size are beyond the reach of what can be achieved on current quantum computers, the possibility of efficiently running large suites of simulations with $\sim 10^{10}$ particles each is still highly valuable to carry out a number of useful calibrations of observational quantities and to explore the parameter space of cosmological models \cite{Castro.etal.2022,Angulo.etal.2021}.
We thus consider a situation of possible cosmological interest to be a $3$D simulation with resolution in grid points per dimension of $2048 = 2^{11}$.
Thanks to the logarithmic encoding, a total of $2^{33}$ grid points can be obtained with $n_{tot} = 33$ qubits.
In Tab.~\ref{tab: different_terms} we report the number of qubits needed for every term of Eq.~\eqref{eq: McLach eq. of motion} and the relative number of different circuits used. 
%%%%%%%     AGGIUNTA TIMESTEP
In this exploratory work we used a heuristic number of parameters $M_p$ and timesteps $N_t$ for our simulation. Thus we are not in position of providing an accurate estimate of the number of parameters, or timesteps, required for a relevant  cosmological simulation. 
What we can say is that, such simulation would require a  maximum of $2n+1$ qubits, used in the evaluation of the potential term.
%%%%%%

The implementation of error mitigation protocols for  near-term hardware experiments with noisy devices does not significantly affect the estimated number of resources (e.g., number of qubits and two-qubit gates). In particular, noise mitigation schemes such as probabilistic error cancellation~\cite{vandenBerg2023} (PEC) and probabilistic error amplification~\cite{Kim2023} (PEA) only require additional single qubit operations to implement Pauli twirling~\cite{vandenBerg2023} (for the conversion of coherent to incoherent noise) and dynamical decoupling, with no effect on the overall resource scaling. On the other hand, a significant increase in the number of measurements is expected for both PEC and PEA approaches.
%%%%%%

%%%  TIMESTEP    %%%%
From Table~\ref{tab: hyperparameters} we can retrieve some useful insights about the required timestep (to ensure numerical stability) and the scaling of the target error with the system size. It is worth mentioning that the following outtakes regard the scenario in which the EOM~\eqref{eq: McLach eq. of motion} is integrated by an explicit first order Euler method.
\\
Firstly, we note that to precisely describe the full Hilbert space, the number of parameters $M_p$ should increase by a factor of two with the addition of each qubit.
Furthermore, increasing spatial resolution (number of qubits) necessitates a higher number of timesteps to maintain the desired level of accuracy in describing the dynamics. This phenomenon is analogous to what occurs in classical numerical integration problems, such as spectral methods or $N$-body simulations. On the other hand, when the fidelity $\mathcal{F}$ is held constant, the expected number of timesteps $N_t$ decreases as the number of variational parameters $M_p$ increases.
This trend can be attributed to the fact that the equation being integrated (Eq. \eqref{eq: McLach eq. of motion}) operates within parameter space, whereas the original dynamics (i.e., the Hamiltonian in Eq. \eqref{eq:Hamiltonian}) is only reflected in the vector term (as per Eq. \eqref{eq: McLach V}). Moreover, the variational approach enables the use of a parameter count smaller than the Hilbert space dimension. Consequently, capturing the same dynamics within a sub-manifold, which offers less flexibility in terms of parameter evolution, requires a finer timestep.

In particular, to span the entire Hilbert space, we would need $M_p=2N$ variational parameters. As $M_p$ deviates from this value, our ability to capture dynamical fluctuations diminishes, necessitating more timesteps to accurately track the wavefunction evolution. This provides an explanation for the lower fidelity values observed in Table \ref{tab: hyperparameters} when a larger number of qubits is employed. In such cases, either $M_p$ or the number of timesteps does not increase in alignment with the scaling necessary to maintain fidelity at a stable level.

It is important to further note that this principle is applicable when $M_p > M_{min}$, where $M_{min}$ represents the minimum number of variational parameters required to reproduce the target function within a specified accuracy. This minimum parameter count can change over the course of the wavefunction evolution based on the complexity.

\begin{figure}
    \centering
    \begin{tikzpicture}
        \node at (0,0) {\includegraphics[width = 0.75 \columnwidth]{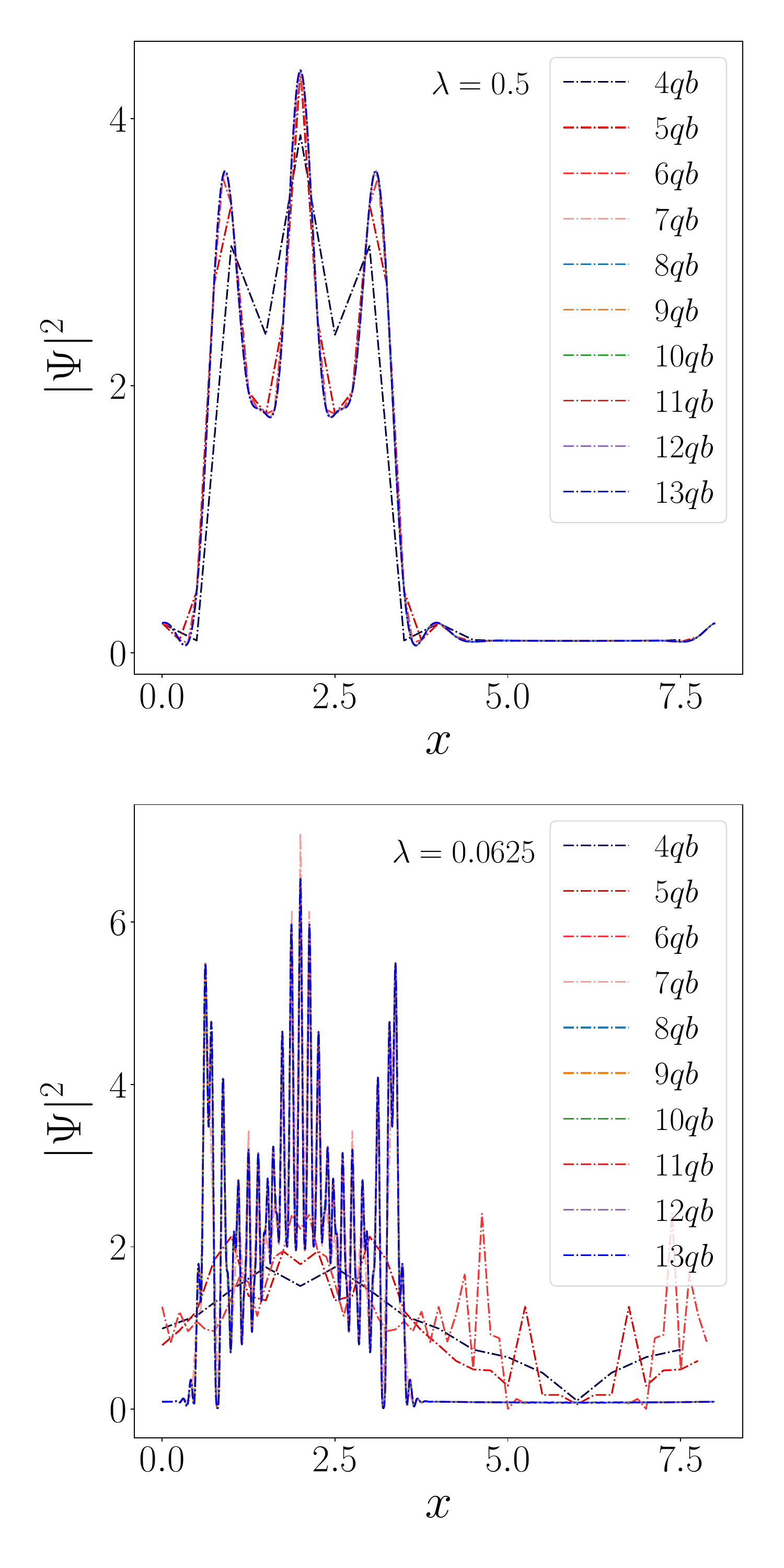}};
        \node at (-3, 5.96) {(a)};
        \node at (-3, -0.35) {(b)};
    \end{tikzpicture}
    \caption{
    Density distribution at a fixed time frame for different resolutions (i.e, number of qubits $n$). On the top (a) and bottom panel (b) the scale $\lambda$ is set respectively to $1/2$ and $1/16$. One can notice how higher resolution is needed to resolve a more classical system (lower $\lambda$).}
    \label{fig: probability_timescreen_different_lambda}
\end{figure}
%---------------------------

\subsubsection{Space resolution and classical limit}\label{classical limit scaling}
%-------------------------
In this preliminary study, we only performed numerical tests on relatively small-scale systems for which numerical simulations of our quantum algorithm were possible with the available computational resources. However, it is essential for us to confirm that the resolution we employed is sufficient to accurately capture the dynamics of the system.
%----------------------
As we approach the classical limit, however, the space resolution needed to capture the right dynamical behaviour increases. 
This is clear in the left panel of Fig.~\ref{fig: scaling_lambda}, where the convergence of the probability distribution is shown as a function of the spatial resolution for simulations with different scale $\lambda$.
We observed that with decreasing $\lambda$ accurate results require a finer representation of the space coordinate. This is mainly due to the appearance of peaked structures observed in the dynamics (see Fig.~\ref{fig: different lambda plots}), which are harder to resolve than in the case of larger $\lambda$ values.
It is worth mentioning that the increase in space resolution also requires a corresponding decrease of the simulation time step (Table~\ref{tab: hyperparameters}). 
\\
In the right panel of Fig.~\ref{fig: scaling_lambda} the resolution is shown as a function of the scale $\lambda$ for different convergence values.
From an empirical fit we showed that the number of qubits necessary to resolve the dynamics of a system scales as $\mathcal{O}(\log(\lambda))$. 
To quantify convergence we used the $L_2$ norm between the $n$ qubits probability distribution $f_n$ -- at a fixed time frame -- mapped onto the $13$ qubits grid and the $13$ qubits probability distribution $f_{n}$
\begin{equation}
    \label{def_mathcalC}
\mathcal{C}^{(13)}_n = || f_{13} - f_n ||_{L_2} \, .
\end{equation} 
In detail, the scaling law is fitted with a logarithmic function $n(\lambda, \tilde{\mathcal{C}}^{(13)}) = K log(\lambda) + q(\tilde{\mathcal{C}}^{(13)}) $, 
where $K = -1.44$ and $q(\tilde{\mathcal{C}}^{(13)})$ is the resolution needed to obtain the desired convergence factor $\tilde{\mathcal{C}}^{(13)}$ when $\lambda = 1$. Here $\tilde{\mathcal{C}}^{(13)}$ indicates a reference value of $\mathcal{C}^{(13)}_n$, chosen a priori, thus does not depends on $n$.

To be able to determine from a qualitative standpoint what value of $\mathcal{C}_{13}$ is needed to obtain convergence in resolution, we plotted in Fig.~\ref{fig: probability_timescreen_different_lambda} the probability distribution at a fixed timestep, for different resolution and different $\lambda$. Comparing the images of this plot with the graphics in Fig.~\ref{fig: scaling_lambda} tells us what \textit{convergence level} is associated to a numerical value of $\mathcal{C}_{13}$. We observed that the right behaviour can be captured as soon as the various density distributions start overlapping. More precisely this happens for $6$ qubits when $\lambda=0.5$ and for $8\div 9$ qubits when $\lambda=0.0.625$. It is fair to assume that a $L_2$ distance of $\mathcal{O}(10^{-1})$ is enough to resolve the dynamic.
We hence gather from both the fit and the previous remarks that a one dimensional resolution of $11$ qubits can be enough to resolve a simulation approaching the classical limit with 
$\lambda$ up to $\mathcal{O}(10^{-3})$.
Moreover, it is possible to show that for the simulations reported in Figure~\ref{fig: 5qb_VTE} with $\lambda =1$ a resolution of $32$ grid points (equivalent to $5$ qubits) is sufficient to converge the primary features of the dynamics. 

%%%%%%%%%%%%%%%%%%%%%%%%%%%%%%%%%%%

%-------------------------
\subsubsection{Sampling and system size}
Of importance is also the study of the convergence of the results as a function of the number of measurements ($N_s$) needed to accurately evaluate the elements in Eqs.~\eqref{eq: McLach M} and~\eqref{eq: McLach V}. 
Measurements
introduces a statistical noise in the solution of the equation of motion for the propagation of the wavefunction parameters, which has an impact on the overall dynamics. 
Building on~\cite{lubasch_variational_2020} we investigate the aforementioned behaviour in the case of the newly introduced term $\bra{\partial_{\theta_k}\psi}\mathcal{H}\ket{\psi}$.
The potential part is directly proportional to the measurement of the the ancilla qubit $\langle \sigma_V^z \rangle$,
thus the variance of the measurements can be estimated by the following 
\begin{equation}
    \label{eq: potential variance}
    \mathcal{E}_V = \phi_V(n) L \sqrt{ \frac{1 - \langle \sigma_V^z \rangle^2}{N_s} } ,
\end{equation}
 where the value of $\sigma^z_V$ is intended in the limit of ${N_s \to \infty}$ and the norm of the potential $\phi_V(n)$ scales with the number of qubits as $2^{n/2}$ (this can be easily seen applying the spectral method proposed in Ref.~\cite{mocz_towards_2021} to obtain the potential, where the wavefunction is normalized as in Eq.~\eqref{eq:physical_quantum_relation}). The fact that the number of shots scales exponentially with the number of qubits is related to the nonlinear nature of the problem. 
 Precisely, it is a consequence of the factorization of the physical wavefunction and the potential (remember Eqs.~\eqref{eq: generic_wavefunc},~\eqref{eq: potential structure}), where the normalization factor appears as an additional parameter that depend on the number of grid points.
 
 The kinetic part is given by a linear combination of three different set of measurements, see Eq.~(\ref{eq: mixed finite diff kinetic}). The variance is estimated with a quadrature-sum as
 \begin{equation}
     \label{eq:kinetic variance}
     \mathcal{E}_K = \frac{2^{2n}}{L} \sqrt{\frac{4 - \langle \sigma_{k+}^z  \rangle^2 
     -\langle\sigma_{k-}^z\rangle^2 +
     2\langle\sigma_{k}^z\rangle^2
     }{N_s}} .
 \end{equation}
 Here the factor $2^{2n}$ emerges from the term $1/\Delta x^2$ required form the finite differences method.
 We observe that in both situations the number of measurement required for an arbitrary accuracy increases with the number of qubits. 
%------------------------

\section{conclusions}\label{conclusions}
In this paper, we tackled the problem of simulating a many-body problem of collisionless self-gravitating particles interacting only through a potential. In a cosmological context, this describes, e.g., the case of gravitational instability of a cold dark matter fluid in an expanding background. Our analysis builds on the possibility to recover the dynamics of the Vlasov-Poisson (VP) equations by mapping it to a framework more suited for quantum computing (QC), namely the Schr\"odinger-Poisson (SP) equations.

We proposed a variational time-evolution (VTE) algorithm for the solution of the corresponding non-linear time-dependant Schr\"odinger-like equation (TDSE) in which, at each time-step, the potential, which is a functional of the time-evolved system wavefunction, is obtained upon minimization of a suitable parameterized unitary in the quantum register. 

The proposed quantum algorithm was developed with the aim of scaling up to system sizes, which are in principle much less favourable for classical computers than for quantum computers.
To this end, we used a compact (i.e., logarithmic) encoding of the spatial grid (i.e., $n$ qubits describing $2^n$ grid points), while enabling the representation of any self-consistent potential, which can be described by combining a parameterized unitary circuit and classical normalization factors.  
In particular, working with a circuit depth that scales polynomially with the number of qubits, we were able
to reach a final state fidelity of approximately $0.96$ in a 5 qubits simulation.
Concerning the scaling of the VTE circuit, the number of terms required to evolve the wavefunction in a single timestep scales quadratically with the number of variational parameters. 
However, the number of timesteps required to achieve a given fidelity increases as the ratio between the number of variational parameters and the Hilbert space dimension decreases, as shown in Tab.~\ref{tab: hyperparameters}.
This behaviour might be related to the heuristic ansatz used in our implementation (e.g, Figs.~\ref{fig:ry_cnot}, \ref{eq: ansatz_structure}). 
We postpone to future investigations understanding weather using ansatz based on tensor networks (e.g, Matrix Product States), as proposed in~\cite{lubasch_variational_2020, mps_2023}, can bring some improvements. 

In addition, the number of measurements required to reach a desired accuracy shows a polynomial scaling with the number of grid points. We point out that this behaviour is not specifically related to our proposed VTE algorithm, but to the approach chosen to tackle the nonlinear nature of the problem, namely factorizing the potential and the wavefunction into unitary circuits followed by classical normalization.

Moreover, using classical simulations we investigated  how the required resolution changes as we approach the classical limit $\hbar/m \to 0$ in a $1D$ scenario.
The proposed empirical log-scaling law opens up new interesting perspectives for the use of QC in the propagation of the SP equation in more general settings, including the $3D$ case.

In conclusion, we consider this work as a first steps towards the use of QC in the solution of the dynamics of a self-gravitating collisionless fluid.
While the scaling up of the quantum approach to system sizes that may be relevant for cosmological prediction in 3D seams unlikely before the advent of fault-tolerant quantum computing, there may be interesting studies (e.g., the study of static and dynamic phase transitions) which may occur already in low dimensions (1D) and that can become classically hard because of the complexity of the quantum description SP formulation (e.g., because of the growing entanglement). A similar strategy was recently implemented in the domain lattice gauge theory (see~\cite{dimeglio2023quantum}).
It is also worth pointing out that, while this study was inspired by the cosmological problem of gravitational instability of a collisionless fluid, our results are general and can be applied to other domains, including the study of the plasma dynamics in a Tokamak fusion reactor. 

At the current state of development, our QC algorithm is clearly not competitive, in terms of accessible  dynamic range,  with respect to classical  methods, both in cosmology and plasma physics, using near-term, noisy, QC with a number of qubits $\sim 100$ ~\cite{100x100_blog}.  
On the other hand, developments that can make our approach more noise-resilient can still be foreseen, including more efficient integration methods and physically motivated variational ansatz. 

A particularly intriguing prospect is the incorporation of non-variational methods within the time evolution algorithm for solving the Poisson equation (or any other equation where the potential relies on the wavefunction). This approach holds the potential to deliver more precise results at the cost of a significant increase in the circuit depth, likely requiring a fault-tolerant quantum computing implementation~\cite{A_PDE_2017, B_linear_Vlasov, C_efficient_linear_solver} . 
We therefore look with a good deal of optimism into the future developments of this very promising application domain for QC.

\acknowledgements{We thank Guglielmo Mazzola for insightful discussions and feedback 
and the unknown referees for their constructive comments.
This paper is supported by the Fondazione ICSC National Recovery and Resilience Plan (PNRR)
Project ID CN-00000013 "Italian Research Center on High-Performance
Computing, Big Data and Quantum Computing" funded by MUR Missione 4
Componente 2 Investimento 1.4: "Potenziamento strutture di ricerca e
creazione di "campioni nazionali di R$\&$S (M4C2-19)" - Next Generation
EU (NGEU).
We acknowledge the use of IBM Quantum services for this work. IBM, the IBM logo, and ibm.com are trademarks of International Business Machines Corp., registered in many jurisdictions worldwide. Other product and service names might be trademarks of IBM or other companies. 
The current list of IBM trademarks is available at \url{https://www.ibm.com/legal/copytrade.}
}

%%%%%%%%%%%%%%%%%%%%%%%%%%%%%%%%%%%%%%%%%%%%%%%%%%%
%%%%%%%%%%%%%%%%%%%%%%%%%%%%%%%%%%%%%%%%%%%%%%%%%%%
%%%%%%%%%%%%%%%%%%%%%%%%%%%%%%%%%%%%%%%%%%%%%%%%%%%
\newpage
~
\newpage
\appendix
    \section{SP DIMENSION RE-SCALING}\label{supp: SP rescaling}
    Let's consider the Schr\"odinger-Poisson (SP) equation in a format similar to the one presented in~\cite{mocz_schrodinger-poissonvlasov-poisson_2018}:
    \begin{equation}
    \label{eq: supp_starting_SP}
    \begin{aligned}
            &i \hbar \frac{\partial \Psi}{\partial t}=-\frac{\hbar^{2}}{2 m} \nabla^{2} \Psi+m U \Psi\,;
            \\
            &\nabla^{2} U = 4 \pi G(\rho\ - \rho^*)\, ,
    \end{aligned}    
    \end{equation}
    where $\rho^*$ is a reference density. 
    If we chose it to be the average density over the volume, the normalization of the wavefunction as given in the main text directly follows.
    
    We can transform the potential so that
    \begin{equation}
        \label{eq: supp_first_pot_transf}
        \nabla^{2} U = 4 \pi G \rho^*\left(\frac{\rho}{\rho^*} - 1\right )\,.
    \end{equation}
    If we now define $|\Psi|^2 := \rho / \rho^*$ and $\lambda := \hbar / m$, Eqs.\eqref{eq: supp_starting_SP} can be recast in the form
    \begin{equation}
    \label{eq: supp_SP_second}
    \begin{aligned}
            &i \frac{\partial \Psi}{\partial t}=-\frac{\lambda}{2 } \nabla^{2} \Psi + \frac{1}{\lambda} U \Psi\,;
            \\
            &\nabla^{2} U = 4 \pi G\rho^* (|\Psi|^2-1)\,.
    \end{aligned}    
    \end{equation}
    The potential we used in the simulation is redefined so that the Poisson equation is adimensional. This is done defining a function $V:= U / \alpha$, so that
    \begin{equation}
        \nabla^2 V = |\Psi|^2 -1\,,    
    \end{equation}
    with $\alpha = 4 \pi G \rho^*$.  We have now to substitute $U(V)$ in Eq.~\eqref{eq: supp_SP_second}:
    \begin{equation}
        \label{eq: supp_SP_third}
        \begin{aligned}
            &i \frac{\partial \Psi}{\partial t}=-\frac{\lambda}{2} \nabla^{2} \Psi + \frac{\alpha}{\lambda} V \Psi\,;
            \\
            &\nabla^2 V = |\Psi|^2 -1.
        \end{aligned}    
    \end{equation}
    Finally the equation we worked with is obtained if and only if 
    \begin{equation} \label{eq: supp: condition on rho}
        \alpha = 1 \iff \rho^* = \frac{1}{4 \pi G}
        \iff \frac{4\pi G}{L}\int_0^L \rho(x) dx = 1\,,
    \end{equation}
    where we used the definition of $\rho^*$. This sets the computational units as combinations of $[G, \rho^*, L]$. 
    
    Is worth noticing that from a dimensional point of view, the first equation holds only if the density $\rho$ is defined in 3D.
    If this is not the case, a correcting factor is needed. When we work in one dimension, we assume spherical symmetry, so that the dependence of the functions involved is one dimensional, i.e. there is only dependence on a radial coordinate $r$, but the density remains three-dimensional (i.e, $[\rho(r)] = 1 / L^3$).
    %\end{section}
    
    %%%%%%%%%%%%%%%%%%%%%%%%%%%%%%%%%%%%%%%%%%%%%%%%%%%%%%%%%%%%%%%%
    %%%%%%%%%%%%%%%%%%%%%%%%%%%%%%%%%%%%%%%%%%%%%%%%%%%%%%%%%%%%%%%%
    
    \begin{section}{POTENTIAL COST FUNCTION}\label{sec: potential}
    In the main text, the cost function used to find the solution of the Poisson equation,(see Eq.~\eqref{eq:analytic_cost_potential} in main text) is a Euclidean norm of a vector
    \begin{equation}
        \begin{aligned}
            \label{eq: supp_pot1}
            || \nabla^2V(\phi) - |\Psi(\theta)|^2 +1||^2
            = 
            \\
            \sum_{j=0}^{N-1} 
            \left(
            {\nabla^2V_j(\boldsymbol{\phi}) - |\Psi_j(\boldsymbol{\theta})|^2 + 1}
            \right)^2 \, .
        \end{aligned}
        %\left(\sum_k {\nabla^2V_k(\boldsymbol{\phi}) - |\Phi_k(\boldsymbol{\theta})|^2 + 1}\right)
    \end{equation}
    Developing this relation, grouping all terms and taking into consideration only those explicitly dependant on $\boldsymbol{\phi}$ we found
    \begin{equation}
    \begin{aligned}
        \label{eq: supp_pot2}
            min_{\boldsymbol{\phi}} \Big\{  
                |\nabla^2 V(\boldsymbol{\phi})|^2 
                & - 2\Re\{\nabla^2 V (\boldsymbol{\phi}) \cdot |\Psi|^2 \} 
                \\ & 
                %\nonumber
                + 2 \Re\{\nabla^2V (\boldsymbol{\phi}) \cdot \mathbf{1}\}
            \Big\}    \, ,    
    \end{aligned}
    \end{equation}
    where we used for conciseness $\Psi =: \Psi(\boldsymbol{\theta})$.
    \\
    %Remembering $V(\boldsymbol{\phi}) = \phi_V \tilde{V}(\tilde{\boldsymbol{\phi}})$
    Due to the PBC of the problem, the last term of the previous equation vanishes if we use a finite differences approach for the evaluation of the Laplacian
    ${\nabla^2 V_j = (V_{j+1} - 2V_j + V_{j-1}) / \Delta x^2}$.
    Where we omitted the parameters' dependence of $V$ just for brevity.
    \\
    Let's focus in the first term of Eq.~\eqref{eq: supp_pot2}. Remembering the normalization of the potential (Eq.~\eqref{eq: potential structure} in main text) and the PBC one finds that
    %\begin{equation}
    \begin{align}
        \label{eq: supp_pot_cost1}
        &\sum_{j=0}^{N-1}
        \phi_V^2
        \left(
            \frac{\Tilde{V}_{j+1} - 2\Tilde{V}_j + \Tilde{V}_{j-1}}{\Delta x^4}
        \right)^2
        = \\ \nonumber
        & 
        \frac{2\phi_V^2}{\Delta x^4}
        \left[
        4\left(1 - \sum_{j=0}^{N-1} \Tilde{V}_j \Tilde{V}_{j+1} \right) 
        - \left( 1 - \sum_{j=0}^{N-1} \Tilde{V}_j \Tilde{V}_{j+2} \right)
        \right] \, .
    \end{align}
    %\end{equation}
    This term can be computed using a circuit like the one in Fig.~\ref{fig: new mixed potential} in the main text, where instead of $F(\boldsymbol{\theta})$ one uses the unitary ansatz for the potential $U(\boldsymbol{\Tilde{\phi}})$ (Fig.~\ref{fig:ry_cnot} in main text). This is possible because the potential is real valued, and by evaluating $\langle\langle\sigma_z\rangle\rangle$ on the ancilla qubit 
    \begin{equation}
        \label{eq: exp.val.V}
        1 - \langle\sigma_z\rangle =
        1 - \sum_{j=0}^{N-1} V_j V_{j+1}   \, . 
    \end{equation}
    If two adder circuits $A$ are used instead, from the previous relation is possible to retrieve $1 - \sum_{j=0}^{N-1} V_j V_{j+2}$.
    
    Switching to the second term in Eq.~\eqref{eq: supp_pot2} and unraveling the Laplacian we find that three terms need to be evaluated.
    \begin{equation}
    \begin{aligned}
        \label{eq: supp_pot_cost2}
        \sum_{j=0}^{N-1}
        &    \phi_V  |\Psi_j|^2 \frac{V_{j+1} - 2V_j + V_{j-1}}{\Delta x^2} \cdot \Delta x^2
        \\ %\nonumber 
        & = \bra{\Psi}V_+\ket{\Psi} - 2 \bra{\Psi}V\ket{\Psi} + \bra{\Psi}V_-\ket{\Psi} \, ,
    \end{aligned}
    \end{equation}
    where $V_\pm$ are the shifted version of the potential (analogous to the one in Eq.~\eqref{eq: mixed finite diff kinetic} in main text). These three expectation value can be evaluated with a similar circuit to the one in Fig.~\ref{fig: new mixed kinetic} in the main text, where instead of $F(\boldsymbol{\theta})$ one uses the unitary ansatz for the wavefunction $U(\boldsymbol{\theta})$. The shifted potential is obtained with the adder circuit and its inverse.
    \end{section}
    %%%%%
    %%%%%%%%%%%%%%%%%%%%%%%%%%% FINE POTENZIALE     %%%%%%%%%%%%%%%%%%%%%%%%%%%%
    %%%%%%%%%%%%%%%%%%%
    %%%%%%%%%%                              ADDER CIRCUIT     %%%%%%%%%%%%%%%%%%
    \begin{section}{ADDER CIRCUIT}\label{sec: supp: adder}
    This circuit scheme is taken from \cite{lubasch_variational_2020}. Here we present it with some more details about its working principles.\\
    The action of the adder circuit on the $j$-th basis produces a negative shift $A:\ket{bin(j)}\mapsto\ket{bin(j-1)}$. When this is applied to a generic wavefunction it results in a shift of the coefficient that takes into consideration periodic boundary conditions $A\ket{\psi}=\sum_{j=0}^{N-1}\psi_{j+1}\ket{bin(j)}$. In this work, the action of the unitary $A$ is controlled by an ancillary qubit.
    
    The philosophy behind its working principle is that, to produce the desired shift, (\textit{i}) first the least significant (LSQ) qubit have to be negated using one $CX$ gate, (\textit{ii}) then is added to the second LSQ using a Toffoli controlled by the ancilla and the LSQ. 
    (\textit{iii}) Going up in the hierarchy, to the following qubit is added a product of the previous state. This is repeated until the most significant qubitis reached. The product of the previous states is s stored in $n-2$ ancillary qubits. The \textit{loading} process is carried out by n-2 Toffoli gates, while the \textit{adding} uses $n-2$ $CX$ gates and Two Toffoli. 
    (\textit{iv}) To finish the procedure, the ancillary register needs to be set back to the initial state $\ket{\boldsymbol{0}}$ using $n-2$ Toffoli. 
    The implementation of the adder circuit requires a total of $2n-2$ Toffoli, $n-2$ $CX$ gates and $n-2$ ancillary qubits. In Fig.~\ref{fig: Adder} is shown an example for the case of a 4 qubits system.
    \\ The unitary $A^{-1}$ produces a positive shift and is obtained reverting the adder (i.e, from right to left).
    %%% FIGURE 
    \begin{figure}
        \resizebox{\columnwidth}{!}{% code from SebGlav answer
         \begin{quantikz}
             \lstick{$\ket{+}$} &\ctrl{4} &\ctrl{4} &\ctrl{1} &\qw &\qw
             &\qw &\qw 
             &\qw &\qw  &\qw &\ctrl{1}
             \\
             \lstick{$a_0 = \ket{0}$} &\qw &\qw &\gate{X} &\ctrl{1} &\qw
             &\qw &\qw 
             &\qw &\qw  &\ctrl{1} &\gate{X}
             \\
             \lstick{$a_1 = \ket{0}$} &\qw &\qw &\qw &\gate{X} &\ctrl{4}
             &\ctrl{1} &\qw 
             &\qw &\ctrl{1} &\gate{X} &\qw
             \\
             \lstick{$a_2 = \ket{0}$} &\qw &\qw &\qw  
             &\qw &\qw&\gate{X} &\ctrl{4} &\ctrl{4} &\gate{X}
             &\qw &\qw
             \\
             \lstick{$qr_0$} &\gate{X} &\ctrl{1} &\ctrl{-3} &\qw &\qw 
             &\qw &\qw 
             &\qw &\qw  &\qw &\ctrl{-3}
             \\
             \lstick{$qr_1$} &\qw &\gate{X} &\qw &\ctrl{-3} &\qw 
             &\qw &\qw 
             &\qw &\qw &\ctrl{-3} &\qw
             \\
             \lstick{$qr_2$} &\qw &\qw &\qw &\qw &\gate{X} 
             &\ctrl{-3} &\qw 
             &\qw &\ctrl{-3}  &\qw  &\qw
             \\
             \lstick{$qr_3$} &\qw &\qw &\qw &\qw &\qw 
             &\qw &\gate{X} &\ctrl{1} &\qw &\qw &\qw 
             \\
             \lstick{$qr_4$} &\qw &\qw &\qw &\qw &\qw 
             &\qw &\qw &\gate{X} &\qw &\qw  &\qw 
        \end{quantikz} 
        }        
        \caption{ Example of an adder for a 5-qubits system. The first qubit on the top is the control. The most significant qubit is the one at the bottom.}
        \label{fig: Adder}
    \end{figure}
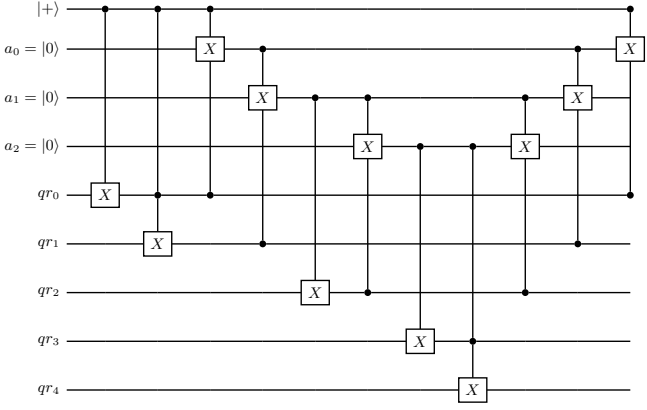
    \end{section}
    
    %%%%%%%      SCHEMA ALGORITMO      %%%%%%%%
    
    %%%%%%%%%%%%%%%%%%%%%%%%%%%%%%%%%%%%%     FINE ADDER    %%%%%%%%%%%%%%%%%
    %%%%%%%%%%%%%%%%%%%%%%%%
    
    %%%%%%%%%%%%%%%%%%%%%%%%%%%%%%%%%%%%      INIZIO Proofs     %%%%%%%
    \begin{section}{CIRCUIT PROOFS}\label{sec: supp: circuit proofs}
    This section gives an idea on how the circuits proposed in Fig.~\ref{fig: new circuit hamiltonian} of the main text work.
    \begin{subsubsection}{Potential}  
    Consider the circuit used for the evaluation of $\bra{\partial_{\theta_k}\psi} \Tilde{V}\ket{\psi}$.
    We refer to the the quantum state before the application of the Toffoli ladder with
    \begin{equation}
        \label{eq: app: pot_1st output}
        \ket{\Xi}_0 = 
        \frac{1}{\sqrt{2}} \left( 
        2i\ket{\boldsymbol{0}} \ket{\partial_{\theta_k}\psi} \ket{0} + \ket{\Phi_{\Tilde{V}}} \ket{\psi} \ket{1}
        \right),
    \end{equation}
    where we used Eq.~\eqref{eq: structure of states for mixed} of the main text and the fact that we can encode the wavefunction and the potential using parameterized circuits. If we omit the explicit parameters' dependence we have that the wavefunction can be encoded as
    \begin{equation}
        \label{eq: app: wavefunc encoding}
        \ket{\psi} = \sum_{j=0}^{N-1} \psi_j \ket{bin(j)} \, ,
    \end{equation}
    and the potential
    \begin{equation}
        \label{eq: app: wavefunc encoding2}
        \ket{\Phi_{\Tilde{V}}} = \sum_{j=0}^{N-1} \Tilde{V}_j \ket{bin(j)} \, .
    \end{equation}
    With $bin(j)$ we refer to the binary conversion of the decimal number $j$. The Toffoli gate adds to the control qubit the product of the two control states.
    \begin{equation}
        \label{eq: app: pot_1_dopo_toffoli}
        \begin{aligned}
        \ket{\Xi}_1 = &
        \frac{1}{\sqrt{2}} \Big( 
        2i\ket{\boldsymbol{0}} \ket{\partial_{\theta_k}\psi} \ket{0} + 
        \\ &
        \sum_{j,l = 0}^{N-1} \psi_l \Tilde{V}_{\Tilde{J}(j,l)} \ket{bin(j)} \ket{bin(l)} \ket{1}
        \Big) \,,
        \end{aligned}
    \end{equation}
    where we defined $\Tilde{J}(j,l) = dec(bin(j) + bin(l))$ as the decimal conversion of the binary sum of the indices $j$ and $l$, with periodic boundary conditions (e,g. with $N = 4$, $\Tilde{J}(1,3) = dec(01 + 11) =  dec(00) = 0$; with $N = 8$, $\Tilde{J}(1,3) = dec(001 + 011) =  dec(100) = 4$).
    We point out that $\Tilde{J}(0,l) = l$. With this relation in mind one can write the quantum state before the measurement as
    \begin{equation}
        \label{eq: app: pot_before measure}
        \begin{aligned}
        \ket{\Xi}_2 = &
        \frac{1}{2}
        \sum_{l = 0}^{N-1}
        \Big[
        \left(  2i\partial_k\psi_l \pm \Tilde{V}_l\psi_l) \ket{\boldsymbol{0}}  \right)
        \\ &
        \pm
        \sum_{j=1}^{N-1} \psi_l \Tilde{V}_{\Tilde{J}(j,l)} \ket{bin(j)}
        \Big]
        \ket{bin(l)} \ket{0/1}
         \,,
        \end{aligned}
    \end{equation}
    where the sign $+$ (or $-$) is used when the ancilla qubit in the state $\ket{0}$ (or $\ket{1}$).
    \\
    Evaluating $\sigma^z$ with this quantum state is equivalent to find the probability of having outcome $0$ minus the one of outcome $1$.
    \begin{equation}
        \label{eq: app: Prob pot}
        \begin{aligned}
            P(0/1) = &
            \frac{1}{4} \sum_{l = 0}^{N-1} \Big[
                4|\partial_k\psi_l|^2 +  |\psi_l|^2
                \big( \Tilde{V}_l^2 + \sum_{j =1}^{N-1} \Tilde{V}_{\Tilde{J}(j,l)}^2 \big)
                \\ &
                \mp 2i  \left( \partial_k\psi_l^* \Tilde{V}_l \psi_l - \partial_k\psi_l \Tilde{V}_l \psi_l^* \right) 
            \Big] \, .
        \end{aligned}
    \end{equation}
    We observe that $\Tilde{V}^2_l$ corresponds to $\Tilde{V}^2_{\Tilde{J}(j=0,l)}$; this way 
    $\Tilde{V}_l^2 + \sum_{j =1}^{N-1} \Tilde{V}_{\Tilde{J}(j,l)}^2 = \sum_{j = 0}^{N-1} \Tilde{V}_{\Tilde{J}(j,l)}^2$, where the periodic boundary conditions assure us that, for a given $l$, this is equivalent to the squared module of the quantum state $\ket{\Phi_{\Tilde{V}}}$.
    Remembering now the normalization of quantum states we can write 
    \begin{equation}
        \label{eq: app: pot sigma z}
        \begin{aligned}
        \langle \sigma_V^z \rangle 
        & = P(0) - P(1) 
        \\ & = 
        -i \sum_{l=0}^{N-1}(\partial_k\psi_l^* \Tilde{V}_l \psi_l - \partial_k\psi_l \Tilde{V}_l \psi_l^*)  
        %\\ &
        %= -i (A - \Bar{A}) = -i (2i \Im A) = 2\Im A = 
        \\ & = 2 \Im \left\{ \sum_{l=0}^{N-1} \partial_k\psi_l^* \Tilde{V}_l \psi_l  \right\} \, .
        \end{aligned}
    \end{equation}
    \end{subsubsection}
    %%%%%%%%%%%%%%%%%%%%%%%%%%%%%%%%%%%% CINETICA  %%%%%%%%%%%%%%%%%%%%%%%%%%
    \begin{subsubsection}{Kinetic term}\label{supp: kinetic}
        Consider Eq.~\eqref{eq: mixed finite diff kinetic} of the main text. The evaluation of the overlap $\braket{\partial_{\theta_k}\psi | \psi}$ is easy and has already been tackled in the main text as well as in~\cite{ollitrault_quantum_2022}. This subsection analyzes the implementation of the circuits in Fig.~\ref{fig: new mixed potential} in the main text.
        \\
        Let's consider the case in which the derivative is controlled by the ancilla qubit in the state $\ket{0}$ ($F_k^{(0)}$ is used).
        Since the adder circuit (\ref{sec: supp: adder}) is controlled by the ancilla state $\ket{1}$, the quantum state after its application is 
        \begin{equation}
            \label{eq: app: kin 0}
            \ket{\Xi}_0 = \frac{1}{\sqrt{2}} 
            \sum_{j=0}^{N-1}
            \left( 
            2i \ \partial_k\psi_j \ket{bin(j)}\ket{0}
            + \psi_{j+1}\ket{bin(j)}\ket{1}
            \right) \, .
        \end{equation}
    The quantum state on which $\langle \sigma^z_- \rangle$ is evaluated is given by
    \begin{equation}
            \label{eq: app: kin 1}
            \ket{\Xi}_1 =  \frac{1}{2} 
            \sum_{j=0}^{N-1}
            \left(
            2i \ \partial_k\psi_j \pm \psi_{j+1} 
            \right)
            \ket{bin(j)}\ket{0/1} \, .
        \end{equation}
    Now, in a similar manner to what has been done in the case of the potential is easy to find that
    \begin{equation}
        \label{eq: app: final kin 0}
        \langle \sigma_-^z \rangle = P(0) - P(1) = 2\Im \left\{ \sum_j \partial_k\psi_j^* \psi_{j+1} \right\}
    \end{equation}
    
    Following the same procedure, but in the case when the derivative and the adder are controlled by the same ancilla state (e.g, $\ket{1}$ if using $F_k^{(1)}$) one find that the final state is given by
        \begin{equation}
            \label{eq: app: kin 2}
            \ket{\Xi}_1 =  \frac{1}{2} 
            \sum_{j=0}^{N-1}
            \left(
            \psi_j \pm
            2i \ \partial_k\psi_{j+1} 
            \right)
            \ket{bin(j)}\ket{0/1} \, .
        \end{equation}
    Since the indices follow periodic boundary conditions, what really matter is the relative shift between $\psi$ and $\partial_k\psi$.
    We thus find that
    \begin{equation}
        \label{eq: app: final kin 3}
        \langle \sigma_+^z \rangle = P(0) - P(1) = 2\Im \left\{ \sum_j \partial_k\psi_j^* \psi_{j-1} \right\} \, .
    \end{equation}
    \end{subsubsection}
    \end{section}

%%%%%%%%%%%%%%%%%%%%%%%%%%%%%%%%%%%%%%%%%%%%%%%%%%%

%\bibliography{ref.bib}
%

\end{document}